\documentclass{iopjournal}

\usepackage{amsmath}
\usepackage{amssymb}
\usepackage{centernot}
\usepackage{xfrac}

\DeclareRobustCommand{\divisible}{\mathrel{\vbox{\baselineskip.65ex\lineskiplimit0pt\hbox{.}\hbox{.}\hbox{.}}}}
\DeclareRobustCommand{\ndivisible}{\centernot\divisible}
\newcommand{\defeq}{\stackrel{\text{def}}{=}}
\newcommand{\inb}[3]{{\left#1 #2 \right#3}}
\newcommand{\reff}[1]{{Fig.\ref{fig:#1}}}
\newcommand{\refe}[1]{{Eq.\ref{eq:#1}}}

\newcommand{\bs}[1]{\boldsymbol{#1}}
\newcommand{\squad}{\ \ }
\newcommand{\A}{\text{A}}
\newcommand{\B}{\text{B}}
\newcommand{\C}{\text{C}}
\newcommand{\op}{\text{\tt{(}}}
\newcommand{\cl}{\text{\tt{)}}}
\newcommand{\I}{\mathcal{I}}
\newcommand{\h}{\mathcal{H}}
\newcommand{\X}{\mathcal{X}}
\newcommand{\half}{{\scriptstyle\sfrac{1}{2}}}

\usepackage{tikz,scalerel}
\newcommand{\CP}[3]{
  \draw[#1, very thick] (.5-#2,.1)--(.5-#2,0);
  \draw[#1, very thick] (.5+#2,.1)--(.5+#2,0);
  \draw[#1, very thick] (.5-#2,1-#3)--(.5-#2,1);
  \draw[#1, very thick] (.5+#2,1-#3)--(.5+#2,1);
  \draw[#1, very thick] (.5-#2, #3) arc (180:0:#2);
  \draw[#1, very thick] (.5-#2, 1-#3) arc (-180:0:#2);
}
\newcommand{\LP}[2]{
  \draw[#1, very thick] (.5-#2,0)--(.5-#2,1);
  \draw[#1, very thick] (.5+#2,0)--(.5+#2,1);
}
\newcommand{\PAD}[1]{
  \draw[white] (.5,-#1)--(.5,1+#1);
}
\newcommand{\IU}[3][\Bigg(]{
\scalerel*{\begin{tikzpicture}[baseline=(current bounding box.center)]
  \PAD{.2} \LP{#2}{.43} \CP{#3}{.3}{.1}
\end{tikzpicture}}{#1}}
\newcommand{\UI}[3][\Bigg(]{
\scalerel*{\begin{tikzpicture}[baseline=(current bounding box.center)]
  \PAD{.2} \CP{#2}{.43}{0} \LP{#3}{.3}
  \draw[black, line width=2.5pt] (.1,-.1)--(.9,1.1);
\end{tikzpicture}}{#1}}
\newcommand{\II}[3][\Bigg(]{
\scalerel*{\begin{tikzpicture}[baseline=(current bounding box.center)]
  \PAD{.2} \LP{#2}{.43} \LP{#3}{.3}
\end{tikzpicture}}{#1}}
\newcommand{\UU}[3][\Bigg(]{
\scalerel*{\begin{tikzpicture}[baseline=(current bounding box.center)]
  \PAD{.2} \CP{#2}{.43}{.0} \CP{#3}{.3}{.0}
\end{tikzpicture}}{#1}}

\begin{document}

\articletype{Paper} 

\title{The $\mathbb{Z}_N^{\times 3}$ symmetry protected boundary modes
in two-dimensional Potts paramagnets}

\author{Hrant Topchyan$^{1,*}$\orcid{0000-0001-6269-0236}}

\affil{$^1$A.Alikhanyan National Science Laboratory (Yerevan Physics Institute), Yerevan, Armenia}

\email{hrant.topchyan@yerphi.am}


\begin{abstract}
We construct and analyze a class of one-dimensional boundary Hamiltonians
arising from two-dimensional symmetry-protected topological phases with
$\mathbb{Z}_N^{\times 3}$ symmetry on a triangular lattice.
Using a cohomology-based transformation, the lattice models for the boundary modes
are explicitly obtained, and their structure is shown to be governed by
the arithmetic properties of $N$.
For prime $N$, the boundary theory admits a formulation in terms of
mutually commuting Temperley-Lieb algebras.
For the composite values of $N$, the models exhibit hierarchical or factorized structures.
We demonstrate that the phases can be understood in terms of primary models
augmented by local defect degrees of freedom
that partition the system into independent segments.
Finally, the global symmetry is realized on the boundary
in a non-on-site and anomalous manner,
directly realizing the corresponding 't Hooft anomaly.
\end{abstract}

\newpage

\section{Introduction}

Symmetry-protected topological (SPT) phases have emerged \cite{spt-orig-1,spt-orig-2,spt-orig-3}
as an important paradigm in the study of quantum matter.
They describe gapped systems that exhibit topological features
only in the presence of a ``protecting" symmetry \cite{spt-orig-2,spt-rev-1}.
A distinguishing feature of SPT phases is that
they do not exhibit intrinsic topological order in the bulk
as they are short-range entangled \cite{spt-orig-1, spt-rev-1},
unlike regular topological order \cite{topol-1, topol-2} which is long-range entangled.
SPT phases are characterized by robust boundary phenomena,
including protected gapless modes \cite{spt-orig-2,spt-rev-1} and
anomalous symmetry realizations
\cite{spt-rev-1, thooft-gen-1, thooft-gen-2, spt-cohom-thooft-finite}.
This set of properties makes them promising platforms
for measurement-based quantum computation and potentially fault-tolerant qubits
\cite{pract-1,pract-2,pract-3,pract-z23,pract-zn3}.
A systematic classification of SPT phases in terms of group cohomology
has provided a general framework for understanding these systems
\cite{spt-cohom-1,spt-cohom-2,spt-cohom-3,spt-cohom-thooft-finite}
which has since been extended by subsequent developments
uncovering phases beyond this approach \cite{spt-beyond-1,spt-beyond-2,spt-beyond-3}.

A key feature of the SPT phases is the ’t Hooft anomaly on the boundary of the system, manifested as an obstruction to the realization of the symmetry as
a local associative representation within the Hilbert space of the boundary,
independently of the bulk
\cite{thooft-gen-1,thooft-gen-2,thooft-finite,thooft-twisted,spt-cohom-thooft-finite}.
This anomaly ensures that the boundary cannot be gapped without
breaking the symmetry or introducing additional degrees of freedom \cite{thooft-gen-2}.
As a consequence, the study of boundary theories
is important in understanding the physical content of SPT phases.

Although the group cohomology concept provides an abstract classification of SPT phases,
there are relatively few explicit lattice realizations and studies of the boundary theories
\cite{edge-levin_gu,edge-z33,edge-z3,edge-z2f,edge-z43}.
In particular, the construction of lattice models
with an explicitly written boundary Hamiltonian and its direct analysis
may reveal additional algebraic structures, symmetries,
possible connections to integrable systems and
the conformal field theories corresponding to the continuum limit of the boundary model.

In this work, we consider a family of two-dimensional SPT phases protected by
a $\mathbb{Z}_N^{\times 3}$ symmetry, constructed over a triangular lattice.
Focusing on a specific nontrivial cohomology class that involves all symmetry sectors,
we derive the corresponding one-dimensional boundary Hamiltonians
using a previously developed construction scheme based on a
symmetrized unitary transformation defined through
a nontrivial 3-cocycle of the cohomology group
\cite{spt-cohom-2,spt-cohom-3,edge-levin_gu,edge-z2f}.
The resulting boundary theories are expressed in terms of
constrained $\mathbb{Z}_N$ degrees of freedom with nontrivial interaction rules.

We show that the structure of the boundary Hamiltonian strongly depends
on the arithmetic properties of $N$.
For prime values $f$ of $N$, the boundary theory significantly simplifies
and admits a formulation in terms of local projectors with an enhanced permutation symmetry.
For prime power values $f^\beta$,
the system decomposes into multiple coupled $\mathbb{Z}_f$ sectors,
leading to a hierarchical structure of constraints.
In the most general case of a composite $N$,
the theory factorizes into components associated with the decomposition of $N$ into primes.

A central result of this paper is that all nontrivial SPT boundary theories in this family
can be reduced to a set of primary models supplemented by local ``defect" degrees of freedom.
These defects act as dynamical constraints that split the system into independent segments.
This provides a unified description for the phases in terms of
a smaller set of fundamental building blocks.

For the case of prime $N$, we further demonstrate that the boundary Hamiltonian
has an alternative representation in terms of
two mutually commuting Temperley–Lieb algebras \cite{TL-orig}.
This structure suggests a connection to a class of exactly solvable models
(integrable systems, loop models, statistical mechanics systems)
and provides an analytical route to relate the boundary continuum limit
to the framework of conformal field theories \cite{cft-1,cft-2}.
In addition, boundary theories exhibit an extensive set of conserved quantities,
including winding- and laterality-type charges \cite{edge-z33,edge-z3},
reflecting the constrained nature of the dynamics.

Finally, we analyze the implementation of the global $\mathbb{Z}_N^{\times 3}$ symmetry
on the boundary and explicitly show that it is non-on-site and has an anomalous nature.
By restricting the symmetry to an open segment \cite{thooft-finite,spt-cohom-thooft-finite},
we obtain a representation with a nontrivial associativity phase
which directly reproduces the underlying group cohomology 3-cocycle.
This provides a concrete lattice realization of the ’t Hooft anomaly
associated with the bulk SPT phase.

\section{The considered system}

We start with the $N$-state Potts paramagnet in the non-interacting limit on a
two-dimensional triangular lattice which is given by the Hamiltonian
\begin{equation}
	H_0^N = -\sum_x \sum_{c=1}^{N-1} X_x^c \squad.
    \label{ham_0}
\end{equation}
Here, $X_x$ is the generator of the $\mathbb{Z}_N$ group.
We will work in the basis of operators $n_x$ with
eigenvalues $\{0, 1, \dots, N-1\}$ or an equivalent set of operators
$Z_x = \varepsilon^{n_x}$ with $\varepsilon = \exp (2\pi i/N)$.
$X_x$ is such that
\begin{equation}
    X_x Z_x = \varepsilon Z_x X_x \squad,\squad
    [X_x, Z_{x'}] = 0~\text{ if }~x\neq x'\squad.
\end{equation}
Note that $X_x^N = Z_x^N = \mathbb{I}$.

Similarly to what was done in \cite{edge-z43},
one can define a triangular superlattice over the original lattice,
where each supernode contains three of the original nodes
as shown in \reff{tritri}.
The superlattice has an on-site symmetry of $\mathbb{Z}_N^{\times 3}$.
The state of the supernode $x$ is now defined by values $n_x^\alpha$
with $\alpha \in \{\A,\B,\C\}$ indicating the flavor/color
of the original node within the supernode.

By switching from the original nodes to the supernodes,
the initial on-site symmetry $\mathbb{Z}_N$ is replaced by $\mathbb{Z}_N^{\times 3}$.
Potential applications of systems with this symmetry have
been explored previously \cite{pract-z23,pract-zn3}.

\begin{figure}
	\centering
	\includegraphics[width=.6\linewidth]{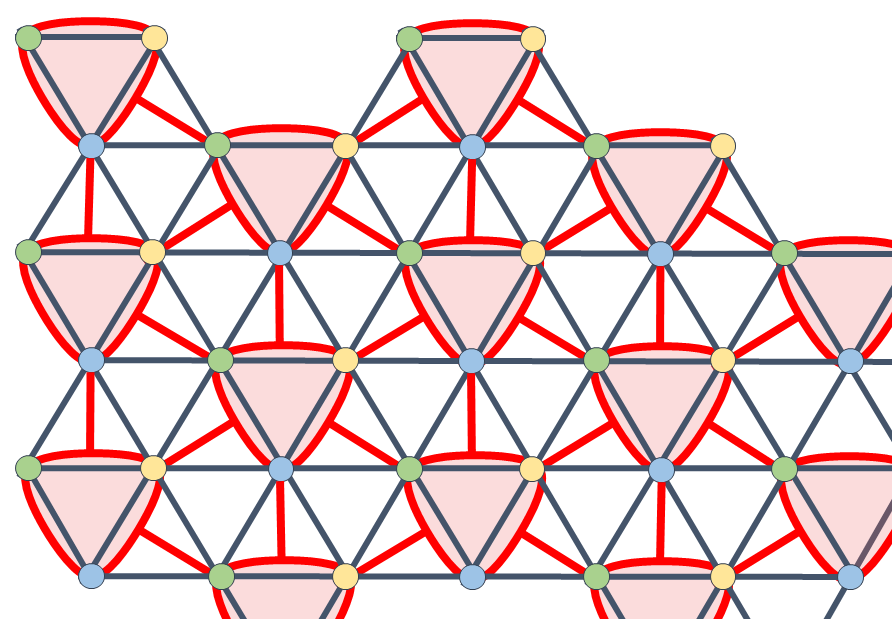}
	\caption{The defined superlattice.
        Green, blue and yellow circles are the nodes of
        the original triangular lattice (shown in dark lines)
        with a $\mathbb{Z}_N$ degree of freedom.
        Red shaded groups are the supernodes with
        a $\mathbb{Z}_N^{\times3}$ degree of freedom,
        that define the triangular superlattice (red lines connecting the supernodes).}
	\label{fig:tritri}
    \vspace{-.5cm}
\end{figure}

\subsection{SPT phases}

The SPT phase space in two dimensions is known to be described by
the cohomologies $H^3(S, U(1))$ of the symmetry group $S$.
\cite{spt-cohom-1,spt-cohom-2,spt-cohom-3,spt-cohom-thooft-finite}.
In our case $S=\mathbb{Z}_N^{\times 3}$, and the relevant cohomology group is
$H^3(\mathbb{Z}_N^{\times 3}, U(1))= \mathbb{Z}_N^{\times 7}$ \cite{cohoms-math},
which can be decomposed as $\mathbb{Z}_N^{\times 3} \times \mathbb{Z}_N^{\times 3} \times \mathbb{Z}_N^{\times 1}$.
The first $\mathbb{Z}_N^{\times 3}$ corresponds to
the three independent $\mathbb{Z}_N$ phases defined on each of the colors,
the second $\mathbb{Z}_N^{\times 3}$ denotes the phases defined on pairs of colors,
and the last $\mathbb{Z}_N$ is the three-color phase space.
We will be working with the last one,
as it is the only one not reducible to smaller symmetry groups.
The corresponding 3-cocycle $\bar{\omega}_3$ can be given by
\begin{equation}
	\bar{\omega}_3(\bs{n}_1, \bs{n}_2, \bs{n}_3) = \varepsilon^{n_1^\A n_2^\B n_3^\C}\squad.
\end{equation}
The closedness of this $\omega_3$ can be checked directly
by verifying that $\delta \bar{\omega}_3(a,b,c,d) = 1$.
The non-exactness can be checked as follows:
for any $\omega_3 = \delta \omega_2$ one can check that
\begin{equation}
	R[\delta \omega_2 (\bs{n}_1, \bs{n}_2, \bs{n}_3)] = 1 \squad \text{with} \squad
	R[\omega_3(\bs{n}_1, \bs{n}_2, \bs{n}_3)] =
	\prod_{a=0}^2 \frac{\omega_3(\bs{n}_{a}, \bs{n}_{a+1}, \bs{n}_{a+2})}
					   {\omega_3(\bs{n}_{a}, \bs{n}_{a-1}, \bs{n}_{a-2})}
\end{equation}
where the addition to the index of $n$ is meant by $\text{mod } 3$ \cite{edge-z33, edge-z43}.
For our $\bar{\omega}_3$ one can check that
\begin{equation}
	R[\bar{\omega}_3(\bs{n}_1, \bs{n}_2, \bs{n}_3)]\bigg|_{n_i^\alpha = \delta_i^\alpha} = \varepsilon
\end{equation}
for later convenience, we will use another nontrivial cohomology group element,
\begin{equation}
	\omega_3(\bs{n}_1, \bs{n}_2, \bs{n}_3) =
    \varepsilon^{n_1^\A \inb({n_2^\B n_3^\C - n_2^\C n_3^\B})}
    \label{eq:cohom}
\end{equation}
The antisymmetrizing counterpart $\omega_3/\bar{\omega}_3$ added to $\bar{\omega}_3$
belongs to the same cohomology class as $\bar{\omega}_3$,
which puts $\omega_3$ into the same cohomology class as $(\bar{\omega}_3)^2$.
It is verified by
\begin{equation}
	\omega_3 =
	(\bar{\omega}_3)^2 \delta \omega_2
	\squad\text{with}\squad
	\omega_2(\bs{n}_1, \bs{n}_2) = \varepsilon^{-n_1^\A n_2^\B n_2^\C} \squad,
\end{equation}
which makes $\omega_3 \equiv (\bar{\omega}_3)^2$.
For even $N$-s, $\omega_3$ is a $\mathbb{Z}_{N/2}$ generator
that no longer generates the initially odd-numbered phases of the $\mathbb{Z}_N$ group.
For odd $N$-s, $\omega_3$ still generates the whole $\mathbb{Z}_N$ group
despite being the square of the original generator.

The alternative (multiplicative) representation $\nu_3$ of the cohomology group generator
that is primarily used in the construction of the boundary modes is given by
\begin{align}
	\nu(0, -\bs{s}, \bs{x}, \bs{y}) =
	\omega_3(&-\bs{s}, \bs{x}+\bs{s}, \bs{y}-\bs{x}) \equiv
	\varepsilon^{\varphi(-\bs{s}, \bs{x}, \bs{y})} \squad,\nonumber\\
	\varphi(-\bs{s}, \bs{x}, \bs{y}) = -s^\A \big(&x^\B y^\C - x^\C y^\B + 
	s^\B \inb({y^\C - x^\C}) - s^\C \inb({y^\B - x^\B})\big) 
\end{align}
 with $\varepsilon = \exp (2\pi i/N)$.
 This form is used in the derivation of nontrivial boundary models
 \cite{edge-z3,edge-z2f}.

\subsection{The boundary modes}

The general procedure to derive the different SPT phase boundary modes for a given symmetry
was developed in \cite{edge-z2f}.
The nontrivial symmetrized Hamiltonian of the $k$-th phase is then given by
\begin{equation}
	H^{N,k} = \frac{1}{|S|}\sum_{\bs{s}} S_{\bs{s}} U_k H_0 U_k^\dagger S_s^{-1}
	\squad,\squad S_{\bs{s}} = \prod_{x, \alpha} X_{x,\alpha}^{\bs{s}^\alpha}
	\squad,\squad U_k = \prod_{\Delta \inb<{pqr}>} \nu^{k\epsilon_\Delta}(0,\bs{n}_x,\bs{n}_q,\bs{n}_r)
    \label{eq:transform}
\end{equation}
where $S_{\bs{s}}$ denotes the global symmetry group operator
with $\bs{s}=\{s^\A,s^\B,s^\C\}$  ($s^\alpha \in \{0, 1, \dots,N-1\}$)
indicating the precise element, and $|S|$ is the size of that group.
$X_{x,\alpha}$ is the operator $\mathbb{Z}_N$ that acts on
the color $\alpha$ of the supernode $x$.
$\epsilon_\Delta$ is a sign factor that
indicates the orientation of the triangle (pointing left or right).
Generally, the nontrivial phases are given by $k\in\{1,2,\dots,N-1\}$,
but for the even $N$-s the relevant values are $k\in\{1,2,\dots,N/2-1\}$,
since $\nu$ is a $\mathbb{Z}_{N/2}$ generator.

The obtained $\varphi$ satisfies the antisymmetry condition
$\varphi(-\bs{s}, \bs{x}, \bs{y})=-\varphi(-\bs{s}, \bs{y}, \bs{x})$,
which allows \cite{edge-z2f} to separate a boundary-shape independent translation-invariant
Hamiltonian
\begin{equation}
    H_{\partial}^{N,k}=-\frac{1}{N^3} \sum_s \sum_{x \in \partial, \alpha} \sum_{c=1}^{N-1}
	V_{\bs{s},x}^k X_{x,\alpha}^c V_{\bs{s},x}^{-k}
    \squad,\squad V_{\bs{s},x} =
	\frac{\nu(0,-\bs{s},\bs{n}_x,\bs{n}_{x-1})}{\nu(0,-\bs{s},\bs{n}_x,\bs{n}_{x+1})}
    \label{eq:V2H}
\end{equation}
for the $k$-th phase's boundary mode.
It is worth mentioning that the operators used in \refe{V2H} are unitarily equivalent
(${\cal O}\rightarrow U_k {\cal O} U_k^\dagger$)
but not exactly the same as the ones used in \refe{transform}.
We will use the notation $-\mathfrak{h}_x^{N,k}$
for the section of $H_\partial^{N,k}$ featuring $X_{x,\alpha}$,
and  $\mathfrak{h}_{x,c}^{N,k}$ for the section of
$\mathfrak{h}_x^{N,k}$ containing a specific order $c$ of $X_{x,\alpha}$,
\begin{equation}
    H_\partial^{N,k}=-\sum_{x\in\partial}\mathfrak{h}_x^{N,k} \squad,\squad
    \mathfrak{h}_x^{N,k} = \sum_{c=1}^{N-1}\mathfrak{h}_{x,c}^{N,k} \squad,\squad
    \mathfrak{h}_{x,c}^{N,k}=\frac{1}{N^3}\sum_{\bs{s},\alpha}
    V_{\bs{s},x}^k X_{x,\alpha}^c V_{\bs{s},x}^{-k} \squad.
\end{equation}

The terms of form $f(\bs{x})-f(\bs{y})$ in $\varphi(-\bs{s},\bs{x},\bs{y})$
do not contribute to the transformation, as they partially vanish in $V_{s,x}$
and the remaining terms commute with $X_{x,\alpha}^c$.
Therefore, we can replace
$\varphi(-\bs{s},\bs{x},\bs{y})\rightarrow -s^\A \inb({x^\B y^\C - x^\C y^\B})$.

Using the relations $X_{x,\alpha}^c \varepsilon^{m n_x^\beta} =
\varepsilon^{m(n_x^\beta + c \delta_\alpha^\beta )} X_{x,\alpha}^c$
for any $m\in\mathbb{N}$ we get
\begin{equation}
    \mathfrak{h}_{x,c}^{N,k}= \frac{1}{N}
	\sum_{s=0}^{N-1}\Big[
		X_{x,\A}^c + Z_{x-1,\C}^{-cks} X_{x,\B}^c Z_{x+1,\C}^{cks}
		+Z_{x-1,\B}^{-cks} X_{x,\C}^c Z_{x+1,\B}^{cks}
	\Big] \squad.
\end{equation}
It consists of a trivial (color $\A$) and nontrivial (colors $\B$ and $\C$) sectors.
The latter is two copies of the same Hamiltonian,
with colors $\B$ and $\C$ in a staggered configuration:
the state of color $B$ in position $x$ interacts only with
the states of color $C$ of its neighbors in $x\pm1$ and vice versa.
For even boundary length $L$, the two copies are independent
as there is even-odd site separation.
For odd $L$, the two merge into a single copy of length $2L$.
The resulting chain will then necessarily have an even length in any case.
The boundary Hamiltonian (the nontrivial sector) is now defined by
\begin{equation}
\begin{split}
    \widetilde{\mathfrak{h}}_{x,c}^{N,k} = \frac{1}{N}
	\sum_{s=0}^{N-1} X_x^c \varepsilon^{ck s(n_{x-1} - n_{x+1})}=
	X_x^c \delta_{\Delta n_x}^{N|ck}
\end{split}
\end{equation}
where we used the definition $Z_x=\varepsilon^{n_x}$ and dropped the color indices as there is only one color involved at each site.
$X_x$ and $n_x$ are the $\mathbb{Z}_N$ operators acting on the ``involved" color.
The notation $a|b$ denotes the quotient of $a$ by
the greatest common divisor of $a$ and $b$ ($a|b=a/\gcd(a,b)$),
$\delta^{m}$ is the Kronecker delta by $\text{mod}~ m$, i.e.
$\delta^{m}_n = \delta_{n ~\text{mod}~ m}$
and $\Delta n_x$ means $n_{x+1} - n_{x-1}$.
The tilde on $\widetilde{\mathfrak{h}}$ indicates that it is not strictly equal to
$\mathfrak{h}$, but is only its nontrivial sector.
The term at site $x$ becomes
\begin{equation}
	\widetilde{\mathfrak{h}}_x^{N,k} =
	\sum_{c=1}^{N-1} X_x^c \delta_{\Delta n_x}^{N|ck} \squad.
	\label{eq:ham_init}
\end{equation}
Although compact, this form is difficult to analyze directly.

\section{Case study of different $N$-values}

For different values of $N$ and $k$, the Hamiltonian
takes different forms. We will now analyze them in detail.

\subsection{Prime values}

For the prime values $N=f$ the Hamiltonian described by \refe{ham_init}
simplifies significantly as $\gcd(f,ck)=1$ for any $0<c, k < f$.
This implies $f|ck=f$, and the boundary Hamiltonian term becomes
\begin{equation}
	\widetilde{\mathfrak{h}}_x^{f,k} =
	(\mathbb{F}_x - \mathbb{I}) \delta_{\Delta n_x}
	\squad,\squad \mathbb{F}_x = \sum_{c=0}^{f-1} X_x^c
	\label{eq:ham_prime}
\end{equation}
where $\mathbb{I}$ is the identity operator.
Note that for the only even prime $f=2$ there are no nontrivial values of $k$.

The operator $\mathbb{F}_x$ has $f-1$ eigenvalues $0$
and a single eigenvalue $f$, making $\mathbb{F}_x/f$ a projector.
In the basis where $n_x$ is diagonal,
$\mathbb{F}_x$ is a matrix whose entries are all equal to $1$.
Equivalently, the action of the Hamiltonian can be described as follows:
if the two neighbors of site $x$ are in the same state,
then the state at $x$ changes to any other allowed state.

Note that the Hamiltonian became independent of the phase index $k$.
This means that all nontrivial boundary modes are described by the same Hamiltonian.
For the case $k \neq 0$ the shorter notation $\widetilde{\mathfrak{h}}_x^f$
and $\widetilde{H}_\partial^f$ with the index $k$ dropped can be used.

\subsubsection{Two-point representation}

The expression for \refe{ham_prime} for a given $x$ contains operators at three sites.
However, it can be reformulated to contain only the nearest neighbor interaction.
To this end, we replace the original
$\mathbb{Z}_f$ operators with barred operators that are defined
separately on even and odd sites as
\begin{equation}
\begin{split}
    \bar{Z}_{2x-1} = Z_{2x-1} Z_{2x}^\dagger \squad&,\squad
    \bar{Z}_{2x} = Z_{2x}^\dagger Z_{2x+1} \\
    X_{2x} = \bar{X}_{2x-1}^\dagger \bar{X}_{2x}^\dagger \squad&,\squad
    X_{2x+1} = \bar{X}_{2x} \bar{X}_{2x+1}
\end{split}
\end{equation}
The notation is consistent, as the system is always even-sized.
The periodicity condition $Z_L=Z_0$ is enforced by demanding
$\bar{\cal Z}\defeq\prod_{x \in \partial} \bar{Z}_x^{(-1)^x} = 1$.

Strictly speaking, the introduced transformation,
analogous to the Kramers–Wannier transformation,
is non-invertible and thus is not a duality.
We define the map to the new Hilbert space
according to the transformation $Z\rightarrow\bar{Z}$.
Then two states of the original Hilbert space
equivalent up to a global ``rotation" ${\cal X}\defeq\prod_{x\in\partial} X_x$
are mapped onto the same state of the new Hilbert space.
The original Hilbert space can be decomposed into a direct sum of $f$ sectors
invariant under Hamiltonian action (since $[H_\partial^f,{\cal X}]=0$)
with eigenvalues $\varepsilon^\lambda$ of $\cal X$.
The map between the ${\cal X}=1$ sector of the original Hilbert space
and the $\bar{\cal Z}=1$ sector of the new Hilbert space is then bijective.
The other sectors of the new Hilbert space $\bar{\cal Z}=\varepsilon^\kappa$
correspond to originally imposed twisted boundary conditions
$Z_{L+x}=\varepsilon^\kappa Z_x$.
The remaining sectors of the original Hilbert space ${\cal X}=\varepsilon^\lambda$
correspond to imposing twisted dual boundary conditions
$\bar{X}_{L+x}=\varepsilon^\lambda \bar{X}_x$.
However, all these constraints are topological and do not affect
the local operator algebra. 
In the remainder of this subsection the boundary-related complications are disregarded.

The new operators satisfy the same commutation relations,
$\bar{X}_x \bar{Z}_x = \varepsilon \bar{Z}_x \bar{X}_x$
and $\bar{X}_x \bar{Z}_{x'} =  \bar{Z}_x \bar{X}_{x'}$ for $x \neq x'$.
We then introduce the two-point operators
\begin{equation}
    {\cal F}_x = \sum_{k=0}^{f-1} \bar{X}_{x-1}^k \bar{X}_x^k \squad,\squad
    {\cal D}_x = \frac{1}{f}\sum_{k=0}^{f-1} \bar{Z}_{x-1}^k \bar{Z}_x^{-k}
\end{equation}
where $x$ enumerates the boundary links.
This brings the Hamiltonian component \refe{ham_prime} to the form
\begin{equation}
    \widetilde{\mathfrak{h}}_x^f \to
    ({\cal F}_x - \mathbb{I}) \cdot {\cal D}_x
    \label{eq:ham_prime_alt}
\end{equation}
and $\widetilde{H}_\partial^f$ is just a summation over all boundary links.
This kind of representation is very useful,
particularly in an attempt to calculate the continuum limit of the theory \cite{continuum},
which is a subject of further study.

The algebraic relations of the introduced operators
can be directly derived from their definitions and are
\begin{equation}
\begin{split}
    {\cal F}_x {\cal F}_x = f {\cal F}_x &\squad,\squad
    {\cal D}_x {\cal D}_x = {\cal D}_x \squad,\\
    [{\cal F}_x, {\cal F}_{x'}]&=
    [{\cal D}_x, {\cal D}_{x'}]=0  \squad,\\
    [{\cal F}_x,{\cal D}_{x'}]=0 &\squad\text{if}\squad
    |x-x'|\neq1 \squad,\\
    {\cal F}_x {\cal D}_{x\pm1} {\cal F}_x = {\cal F}_x &\squad,\squad
    {\cal D}_x {\cal F}_{x\pm1} {\cal D}_x = {\cal D}_x \squad.
\end{split}
\end{equation}
An additional rescaling of $\cal F$ and $\cal D$ as $\check{\cal F}={\cal F}/\sqrt{f}$
and $\check{\cal D}={\cal D}\sqrt{f}$, and
then swapping the definitions of $\check{\cal F}$ and $\check{\cal D}$ on the even links
modifies the algebra to
\begin{align}
    \check{\cal F}_x \check{\cal F}_x= \sqrt{f}\ \check{\cal F}_x  &\squad,\squad\nonumber
    \check{\cal D}_x \check{\cal D}_x= \sqrt{f}\ \check{\cal D}_x  \squad,\\
    [\check{\cal F}_x, \check{\cal D}_{x'}]&=0  \squad,\\
    [\check{\cal F}_x,\check{\cal F}_{x'}]&=
    [\check{\cal D}_x,\check{\cal D}_{x'}]=0  \nonumber
    \squad\text{if}\squad |x-x'|\neq1 \squad, \\
    \check{\cal F}_x \check{\cal F}_{x\pm1} \check{\cal F}_x = \check{\cal F}_x
    &\squad,\squad  \nonumber
    \check{\cal D}_x \check{\cal D}_{x\pm1} \check{\cal D}_x = \check{\cal D}_x
\end{align}
which constitute two mutually commuting Temperley-Lieb (TL) algebras.

The boundary mode is then given by the staggered Hamiltonian
\begin{equation}
    \widetilde{H}_\partial^{f}=-\sum_{x\in\partial} \check{\cal F}_x \check{\cal D}_x +
    \frac{1}{\sqrt{f}} \inb[{\sum_{x\in\partial_\text{e}} \check{\cal F}_x + \sum_{x\in\partial_\text{o}} \check{\cal D}_x}]
\end{equation}
where $\partial_\text{e/o}$ denote even/odd boundary links.
We now introduce the conventional diagram representation of the TL algebra \cite{TL-orig},
depicted in different colors for $\check{\cal F}$ and $\check{\cal D}$.
The following notations are introduced:
\begin{table}[h]
\centering
\setlength{\tabcolsep}{10pt}
\begin{tabular}{c|c|c|c|c}
$x$ & $\mathbb{I}_x$ & $\check{\cal F}_x$ & $\check{\cal D}_x$ &
$\check{\cal F}_x \check{\cal D}_x$ \\ \hline
even & \II{red}{blue} & \IU{red}{blue} & \UI{red}{blue} & \UU{red}{blue} \\
odd & \II{blue}{red} & \UI{blue}{red} & \IU{blue}{red} & \UU{blue}{red} \\
\hline
unified & \II{black}{black} & \multicolumn{2}{c|}{\IU{black}{black}} & \UU{black}{black}
\end{tabular}
\vspace{-.5cm}
\end{table}

\noindent where blue lines are used for the TL algebra of $\check{\cal F}$-s
and red lines for $\check{\cal D}$-s.
The crossed out diagrams emphasize that those specific notations are unnecessary
since the corresponding operators do not appear in the Hamiltonian.
The distinction between the even and odd link definitions is meant to prevent
crossings of blue and red lines in composite diagrams.
All loop weights are $\sqrt{f}$.

The property that lines of different colors
(corresponding to different TL algebras) do not intersect
makes the introduced diagrams ``solid" building blocks for composite diagrams
(composite diagrams are made by stacking the diagrams on top of each other,
preserving the continuity of lines).
We can now introduce colorless (black) diagrams that unify:
the even and odd $x$ diagrams of $\mathbb{I}_x$;
the even and odd $x$ diagrams of ${\cal F}_x{\cal D}_x$;
the even $x$ diagram of ${\cal F}_x$ and the odd $x$ diagram of ${\cal D}_x$.
In the described diagram formulation, the Hamiltonian can be represented as
\begin{equation}
    \widetilde{H}_\partial^{f}=-\sum_{x\in\partial}
    \inb[{\UU[\bigg(]{black}{black} -\frac{1}{\sqrt{f}}\ \IU[\bigg(]{black}{black}}]_x \squad.
    \label{eq:ham_TL}
\end{equation}

The TL algebra hints at a possible connection to integrable models.
In fact, the introduced double-line diagrams constitute
a particular case of the Fuss-Catalan algebra \cite{fuss-catalan}.
The integrable models based on this algebra have been studied in \cite{integr}.
For the case of the model given by \refe{ham_TL},
integrability can be achieved by substituting the second term coefficient,
$-1/\sqrt{f} \rightarrow (f-2)/\sqrt{f}$,
implying integrability only for the trivial case of $f=1$.
The corresponding derivation is presented in the Appendix \ref{sec:integr}.

The emergence of the TL algebra is highly significant.
Although not resulting in integrability,
it provides a natural framework for fermionization \cite{TL-loop-ferm}
and establishes connections to loop models \cite{TL-integr-loop, TL-loop-ferm, TL-loop}
through its diagrammatic formulation.
It is also closely related to the statistical mechanics of low-dimensional quantum systems \cite{TL-rsos}.
The precise implications of this representation are subject to further research.

\subsection{Prime power values}

The Hamiltonians in this formulation are more complex for composite $N$-s.
We will consider the case of $N=f^\beta$ where $f$ is a prime first.
The phase index $k$ and the summation index $c$ in \refe{ham_init} can be represented as
$k=af^\lambda$ and $c=bf^\rho$, with $1\leq a,b\ndivisible f$ and
$0\leq\lambda\leq\beta-1$ (or $\beta-1$ excluded if $f=2$).
Using $\lambda\hat+\rho$ for $\min(\lambda+\rho, \beta)$,
$N|ck$ becomes $f^{\beta-(\lambda\hat+\rho)}$
and $\delta^{N|ck}\rightarrow \delta^{f^{\beta-(\lambda\hat+\rho)}}$.
After adapting the summation by $c$ to this new parametrization,
the expression for $\widetilde{\mathfrak{h}}_x^{N,k}$ becomes
\begin{equation}
	\widetilde{\mathfrak{h}}_x^{f^\beta,af^\lambda} =
	\sum_{\rho=0}^{\beta-1} \inb[{
	\sum_{b=0}^{f^{\beta-\rho}} X_x^{bf^\rho}-
	\sum_{b=0}^{f^{\beta-\rho-1}} X_x^{bf^{\rho+1}}
	}] \delta_{\Delta n_x}^{f^{\beta-(\lambda\hat+\rho)}}
	\label{eq:ham_prpow_init}
\end{equation}
The second sum term in the bracket compensates the terms of the first sum
corresponding to $b \divisible f$, which had been unnecessarily added.
One can directly check that
\begin{equation}
	\sum_{b=0}^{f^{\beta-\rho}} X^{bf^\rho} =
	\mathbb{F}^{({f^{\beta-\rho}})} \otimes \mathbb{I}^{(f^\rho)}	\squad,\squad
	\delta_n^{f^{\beta-(\lambda\hat+\rho)}} =
	\mathbb{I}^{(f^{\lambda\hat+\rho})} \otimes \delta_n^{(f^{\beta-(\lambda\hat+\rho)})}
\end{equation}
where $\mathbb{F}$ is a matrix filled with $1$-s,
$\mathbb{I}$ is the identity matrix,
$\delta$ is the Kronecker matrix,
and the upper indices in parentheses indicate the dimensions of the matrices.
All these objects can be further decomposed as
$\mathbb{F}^{(f^\alpha)} = {\mathbb{F}^{(f)}}^{\otimes\alpha}$,
$\mathbb{I}^{(f^\alpha)} = {\mathbb{I}^{(f)}}^{\otimes\alpha}$,
$\delta^{(f^\alpha)} = {\delta^{(f)}}^{\otimes\alpha}$,
which transforms \refe{ham_prpow_init} to
\begin{equation}
	\widetilde{\mathfrak{h}}_x^{f^\beta,af^\lambda} =
	\sum_{\rho=0}^{\beta-1}
	\inb[{
		 \mathbb{F}^{\otimes(\beta-\rho-1)}
		\otimes (\mathbb{F} - \mathbb{I}) \otimes
		 \mathbb{I}^{\otimes\rho}
	}]_x \cdot \inb[{
		\mathbb{I}^{\otimes(\lambda\hat+\rho)} \otimes
		\delta^{\otimes(\beta-(\lambda\hat+\rho))}
	}]_{\Delta n_x}
\label{eq:ham_prpow_inter}
\end{equation}
where the matrices $\mathbb{F}$, $\mathbb{I}$, and $\delta$ all have dimension $f$,
so the upper index $(f)$ is dropped everywhere.

Now we can see that the degree of freedom $\mathbb{Z}_{f^\beta}$ at each site
can be represented as $\beta$ different $\mathbb{Z}_{f}$ components ($\mathbb{Z}_{f}^{\times\beta}$).
The basis eigenstates $\inb|n>$ of the $\mathbb{Z}_{f^\beta}$ operator $n_x$
can be represented in terms of the eigenstates of the $\mathbb{Z}_{f}^{\times\beta}$ operators
as $\inb|n> = \bigotimes_{\mu=1}^{\beta} \inb|{n_\mu}>$, where $n_\mu\in\{0,1,...,f-1\}$.
The $n$ itself can be represented as $n=\sum_{\mu=1}^{\beta}f^{\beta-\mu}n_\mu$
or in an alternative notation (radix-$f$ number) as $n={\overline{n_1n_2...n_\beta}}$.
Each term of the direct product in the first bracket
acts on its own $\mu$-th subspace of $\inb|{n_\mu}>$.
The factorization of the second term implies
the possibility to use the same basis for the second term as well.
Generally speaking, the direct product terms of the second bracket operator should act on
$\Delta n$ decomposition coefficient space $(\Delta n)_\mu$,
$\Delta n = \overline{(\Delta n)_1...(\Delta n)_\beta}$, and
\begin{equation}
	\overline{(\Delta n)_1...(\Delta n)_\beta} =
	n^+-n^- = \overline{n^+_1...n^+_\beta}-\overline{n^-_1...n^-_\beta}
	\neq \overline{(n^+_1-n^-_1)...(n^+_\beta-n^-_\beta)} =
	\overline{\Delta (n_1)...\Delta (n_\beta)}
\end{equation}
where the notations $n^\pm$ denote $n_{x \pm 1}$,
and the differences of $n_\mu$ are meant by $\text{mod}~ f$.
However, the operator itself made the substitution $(\Delta n)_\mu\rightarrow \Delta(n_\mu)$ possible:
the eigenvalue of the operator is $1$ if and only if $(\Delta n)_\mu=0$
for all $\mu>\lambda+\rho$, and is $0$ otherwise.
On the other hand, the condition $(\Delta n)_\mu = 0$ for all $\mu > \lambda+\rho$
is exactly the same as $\Delta (n_\mu)=0$ for all $\mu > \lambda+\rho$.
In other words, the difference of two numbers' ends in $k$ ``digits" $0$
if and only if the last $k$ ``digits" of the two numbers are the same.
Thus, $\overline{(\Delta n)_1...(\Delta n)_\beta} \equiv \overline{\Delta (n_1)...\Delta (n_\beta)}$
in the context of the second bracket operator, which made substitution possible.

Similarly to the case of prime values of $N$,
the Hamiltonian does not explicitly depend on the phase index $k$,
but does so only through $\lambda$,
meaning that many different boundary modes have the same Hamiltonian description.

\subsubsection{The primary phase}

The first sub-case to consider is the class containing the first nontrivial (``primary") phase
$k=1$ ($\lambda=0$).
In this case, the Hamiltonian \refe{ham_prpow_inter}
can be written in terms of the newly formulated basis $n_{\mu,x}$ as
\begin{equation}
	\widetilde{\mathfrak{h}}_x^{f^\beta,a} =
	\sum_{\rho=0}^{\beta-1}
	\Big[
		 \mathbb{F}_1\cdots\mathbb{F}_{\beta-\rho-1}
		(\mathbb{F} - \mathbb{I})_{\beta-\rho} \cdot
		\delta_{\Delta n_{\rho+1}}\cdots\delta_{\Delta n_\beta}
	\Big]_x
\end{equation}
where $\mathbb{F}_{\mu,x}$ and $\delta_{\Delta n_{\mu,x}}$ act on
the $\mu$-th field component, $n_{\mu,x}$.
We will use the field component and the tensor product notations interchangeably.

Using the fact that the boundary is always even-sized,
we can rearrange the order of the $\beta$ different fields for every second site.
Namely,
for even values of $x$ we replace the field indices $\rho \leftrightarrow \beta-\rho-1$.
For a single site $x$ the corresponding $\mathbb{F}_\rho$-s are given
through even site operators and $\delta_{\Delta n_\rho}$-s through odd site operators,
or vice versa.
After also changing the summation index $\rho \to \beta-\rho$,
$\widetilde{\mathfrak{h}}_x^{f^{\beta},a}$ takes the form
\begin{equation}
	\widetilde{\mathfrak{h}}_x^{f^{\beta}} =
	\sum_{\rho=1}^{\beta}
	\Big[
		\mathbb{F}_1 \cdots \mathbb{F}_{\rho-1} (\mathbb{F} - \mathbb{I})_{\rho}
		\cdot \delta_{\Delta n_1} \cdots \delta_{\Delta n_{\rho}}
	\Big]_x
\end{equation}
where we dropped the index $a$ as it does not matter.
Different terms of the sum commute.

We then introduce a complete set of orthogonal projectors $D^{(\rho)}_{\Delta n} =
\delta_{\Delta n_1}\dots\delta_{\Delta n_\rho}\bar\delta_{\Delta n_{\rho+1}}$
for $0\leq\rho<\beta$ and $D^{(\beta)}_{\Delta n} =
\delta_{\Delta n_1}\dots\delta_{\Delta n_\beta}$.
$\bar\delta_{\Delta n}$ denotes $1-\delta_{\Delta n}$.
Similarly to $\delta_{\Delta_n}$, $D^{(\rho)}_{\Delta_n}$
compares $n_{x-1}$ and $n_{x+1}$ and is equal to $1$
when exactly $\rho$ of their leading components coincide.
Considering the projections of $\widetilde{\mathfrak{h}}_x^{f^{\beta}}$ produces
\begin{equation}
    \widetilde{\mathfrak{h}}_x^{f^{\beta}} D^{(\rho)}_{\Delta_{n_x}} =
    (\mathbb{F}^{\otimes\rho}_x -\mathbb{I})D^{(\rho)}_{\Delta_{n_x}}
\end{equation}
where the right tensor products $\otimes \mathbb{I}$ are dropped.
This reduces the expression of the Hamiltonian component to
\begin{equation}
    \widetilde{\mathfrak{h}}_x^{f^{\beta}} = \sum_{\rho=1}^{\beta}
    \inb[{\mathbb{F}^{\otimes\rho}-\mathbb{I}}]_x D^{(\rho)}_{\Delta n_x} \squad.
	\label{eq:ham_prpow_p1}
\end{equation}
The expression becomes the same as \refe{ham_prime}
for the prime values of $N$ (when $\beta=1$).
Now, the sum terms are orthogonal to each other.
In the sector where $D^{(\rho)}_{\Delta n_x}$
has an eigenvalue $1$ for a given $\rho$,
the eigenvalues of $\widetilde{\mathfrak{h}}_x^{f^{\beta}}$ are
$f^\rho -1$ or $-1$. All possible values of $\rho$
define the complete set of eigenvalues of $\widetilde{\mathfrak{h}}_x^{f^{\beta}}$.

\refe{ham_prpow_p1} can be interpreted as follows:
if the two neighbors of site $x$ have the same state up to component $\rho$,
then the state at $x$ up to component $\rho$ changes to any other allowed state.

\subsubsection{The other phases}

After the factorization procedure in \refe{ham_prpow_inter} one can see
that the first $\lambda$ components
are separated from everything else and become on-site massive free fields.
The eigenvalue of each $\mathbb{F}_{\mu}$ is $f \pi_{\mu}$ where $\pi_{\mu}\in\{0,1\}$.
The operators $\mathbb{F}_{\mu}$ with $\mu\leq\lambda$ can be replaced by their eigenvalues.
The terms with $\rho\geq\beta-\lambda$ in \refe{ham_prpow_inter}
no longer contain operators and produce a constant shift of $f^\lambda\pi_x-1$
where $\pi_x=\prod_{\mu=1}^\lambda \pi_{\mu,x}$ which is also $0$ or $1$.
In the remaining terms, the same substitution by the eigenvalues
introduces a factor $f^\lambda\pi_x$ everywhere.

We can re-enumerate the remaining field components from $1$ to $\beta-\lambda$,
and the operator part of \refe{ham_prpow_inter} with given $\beta$ and $\lambda$
will look similar to the case with $\beta'=\beta-\lambda$, $\lambda'=0$.
After the same manipulations as those done for the $\lambda=0$ case,
the whole expression becomes
\begin{equation}
    \widetilde{\mathfrak{h}}_x^{f^{\beta},af^\lambda} =
    f^\lambda\pi_x \widetilde{\mathfrak{h}}_x^{f^{\beta-\lambda}}
    +f^\lambda\pi_x-1 \squad.
    \label{eq:ham_prpow_pa}
\end{equation}

The phases $af^\lambda$ with $N=f^\beta$ start to resemble
the primary phase of case $N'=f^{\beta-\lambda}$
with allowed static defects (sites where $\pi_x=0$) on the chain.
Phases with different $\lambda\neq 0$ but the same $\beta-\lambda$
are also similar to each other.
They differ only in their energy scale and in the size of the defect subspace
(the number of states for which $\pi_x=0$ or $1$ varies with $\lambda$:
$\pi_x=1$ is a singlet and $\pi_x=0$ is $(f^\lambda-1)$-fold degenerate).

The equation also essentially covers the trivial phase
$k=N$ with $\lambda=\beta$ for which $\widetilde{\mathfrak{h}}_x^1=0$
and we are left with $f^\beta\pi_x-1$,
which are precisely the eigenvalues of $\mathbb{F}_x^{\otimes\beta}-\mathbb{I}$.
The degeneracies of the values of $\pi_x$ also match.

The defects $\pi_x=0$ act as splitting points that divide the chain
into multiple segments with fixed boundary conditions.
Introducing a defect is equivalent to removing the positive semidefinite term
$f^\lambda \inb({\widetilde{\mathfrak{h}}_x^{f^{\beta-\lambda}}+1})$
from \refe{ham_prpow_pa} with $\pi_x=1$.
This generally increases the ground-state energy.
The energy might remain unchanged if the ground state is annihilated
by the removed term, but this is not the generic case.
Hence, the low energy sector of these systems will be described by states without defects.

\subsection{General composite values}

For the general case of a composite $N$,
it can be factorized into prime factors $N=f_1^{\beta_1} f_2^{\beta_2} ... f_D^{\beta_D}$.
Then the $\mathbb{Z}_N$ group can be factorized into independent, mutually commuting  fields
$\mathbb{Z}_N = \mathbb{Z}_{f_1^{\beta_1}}\times\dots\times\mathbb{Z}_{f_D^{\beta_D}}$.

Instead of trying to simplify the expression of \refe{ham_init},
in this case it is easier to start from scratch and use the factorization of the group.
It allows splitting the on-site degree of freedom $n_x$ into $D$ independent components
$n_{d,x}$ and considering the transformations $V_{\bs s,x}$ separately for each of them.
$n_{d,x}$-s can be chosen as $n_{d,x}=n_x~\text{mod}~f_d^{\beta_d}$,
which preserves the group structure.

The initial Hamiltonian given by \refe{ham_init} can be rewritten as
\begin{equation}
    H_0^N=\sum_x \inb({\mathbb{I}-\prod_{d=1}^D \mathbb{F}_{d,x}})
\end{equation}
where $\mathbb{F}_{d,x}$ are the operators that act on the
field components $\mathbb{Z}_{f_d^{\beta_d}}$ at site $x$.
The explicit form of the nontrivial cocycle generator
given by \refe{cohom} in this representation is generally
\begin{equation}
    \omega_3(\bs{n}_1, \bs{n}_2, \bs{n}_3) = \prod_{d=1}^{D}
    \big(\omega_3^{(d)}(\bs{n}_{d,1}, \bs{n}_{d,2}, \bs{n}_{d,3})\big)^{u_d} \squad,\squad
    \omega_3^{(d)}(\bs{n}_{d,1}, \bs{n}_{d,2}, \bs{n}_{d,3}) =
    \varepsilon_{d}^{n_{d,1}^A (n_{d,2}^B n_{d,3}^C - n_{d,2}^C n_{d,3}^B) }
\end{equation}
with integer (B\'ezout) coefficients $u_d$ that depend on $N$,
and $\varepsilon_d = \exp\inb({2\pi i / f_d^{\beta_d}})$.
Here, each factor $\omega_3^{(d)}$ is the nontrivial cocycle
of the $N=f_d^{\beta_d}$ case, acting on the subspace $n_{d,x}$.

The existence of $u_d$-s merely rearranges the set of different powers of $\omega_3$.
It is possible to avoid their introduction by choosing
another nontrivial cohomology element $\omega_3'$ as the generator,
\begin{equation}
    \omega_3' = \prod_{d=1}^{D} \omega_3^{(d)} \squad\Rightarrow\squad
    (\omega_3')^k=\prod_{d=1}^{D} \inb({\omega_3^{(d)}})^{k_d} \squad.
\end{equation}
Here, $k_d$ are phase number decomposition coefficients $k_d=k ~\text{mod}~ f_d^{\beta_d}$.
The closedness and non-exactness of $(\omega_3')^k$ is reduced to the
same properties of $\big({\omega_3^{(d)}}\big)^{k_d}$, which have already been verified
(for a given $k\neq 0$, there is always at least one $k_d\neq0$).
As before, the cocycle does not generate all phases for even $N$-s.

After decomposing the cocycle into terms that act on their specific subspace $n_{d,x}$,
deriving the expression of the boundary Hamiltonian is straightforward: each component 
$\mathbb{F}_{d,x}$ is transformed only by the corresponding ${\omega_3^{(d)}}$.
As the transformation of $\mathbb{F}_{d,x}-\mathbb{I}$ is already known
(\refe{ham_prpow_pa}),
the expression of the generic Hamiltonian component is
\begin{equation}
	\widetilde{\mathfrak{h}}_x^{N,k} = \prod_{d=1}^D
    \inb({\widetilde{\mathfrak{h}}_x^{f_d^{\beta_d},k_d}+\mathbb{I}}) - \mathbb{I} \squad.
    \label{eq:ham_gen_inter}
\end{equation}
The product can also be replaced by a tensor product.

The component-specific phase indices can also be expressed as $k_d=a_df_d^{\lambda_d}$
where $a_d\ndivisible f_d$.
The primary phase (when $\text{gcd}(N,k)=1 \Leftrightarrow \lambda_d=0$) is then given by
\begin{equation}
    \widetilde{\mathfrak{h}}_x^{N} = \prod_{d=1}^D
    \inb({\widetilde{\mathfrak{h}}_x^{f_d^{\beta_d}}+\mathbb{I}}) - \mathbb{I} \squad.
    \label{eq:ham_gen_p1}
\end{equation}
The form for all other phases is obtained by substituting \refe{ham_prpow_pa} into
\refe{ham_gen_inter}, and the result is
\begin{equation}
    \widetilde{\mathfrak{h}}_x^{N, k} = f^{(k)} \pi_x \prod_{d=1}^D
    \inb({\widetilde{\mathfrak{h}}_x^{f_d^{\beta_d-\lambda_d}}+\mathbb{I}}) - \mathbb{I} \squad.
\end{equation}
where $f^{(k)}=\prod_d f_d^{\lambda_d}=\text{gcd}(N, k)$ and
$\pi_x=\prod \pi_{d,x}$ is the composite defect field resulting from different components $d$
($\pi_x$ has a singlet eigenvalue $1$ and
a $(\text{gcd}(N, k)-1)$-fold degenerate eigenvalue $0$).
Note that the non-primary phase $k$ for a fixed $N$
is given by the Hamiltonian term of the primary phase
for $N'=N|k$ with added defects described by $\text{gcd}(N,k)$,
\begin{equation}
    \widetilde{\mathfrak{h}}_x^{N, k} = f^{(k)} \pi_x
    \inb({\widetilde{\mathfrak{h}}_x^{N|k}+\mathbb{I}}) - \mathbb{I} \squad.
    \label{eq:ham_gen_pa}
\end{equation}

Similarly to the case with prime power values of $N$,
here as well the defects $\pi_x=0$ have the role of splitting points
which break the chain into independent segments, and
the low-energy physics of the system will be given by defect-free states.
The number of significantly different boundary modes for a given odd $N$ is
$\prod_d(\beta_d+1)-1$, which is the number of different possible values of
$N|k$ other than $1$.
For even values of $N$, the restriction $k<N/2$
eliminates the possibility $N|k=2$ that occurs only for $k=N/2$.
There are no further limitations since $N|k=N|(N-k)$.
Thus, for an even $N$ the number of different boundary modes is $\prod_d(\beta_d+1)-2$.

\hfill

In summary, every boundary mode is described within the following paradigm.
The first subset of possible boundary modes contains
the primary phase boundary modes for each $N$,
described by \refe{ham_prpow_p1} and \refe{ham_gen_p1}.
The other subset contains the ``diluted" version of primary phase boundary modes
that have an additional defect field, which splits the chain into independent segments.
The ``diluted" primary phase boundary modes of a given $N_0$ appear as
non-primary phase boundary modes for $N$-s that are multiples of $N_0$ (\refe{ham_gen_pa}).

\section{Aspects of symmetries}

In this section, we will discuss the symmetries of the boundary modes
for the different values of $N$ that were obtained above,
starting with the more trivial cases of global symmetries.

The Hamiltonian for prime values of $N=f$ given by \refe{ham_prime}
has two distinct permutation symmetries ${\cal S}_e,{\cal S}_o \cong \mathbb{S}_f$ given by
\begin{equation}
	S_\text{e/o}^{(i)} = \prod_{x\in \partial_\text{e/o}} P_x^{(i)}
	\squad,\squad \inb[{{\cal S}_\text{e/o}, \widetilde{H}_\partial^f}] = 0
	\squad,\squad [{\cal S}_\text{e}, {\cal S}_\text{o}] = 0
\end{equation}
where $\partial_\text{e/o}$ denote even/odd boundary sites and
$P_x^{(i)}$ is the $i$-th permutation operator acting on the $\inb|n>$ space at the site $x$.
Moreover, each operator $\mathbb{F}_x$ and $\delta_{\Delta n_x}$ is individually symmetric.

In the case of the primary phase for a prime power $N=f^\beta$ (\refe{ham_prpow_p1}),
each symmetry (${\cal S}_\text{e}$ and ${\cal S}_\text{o}$) is promoted to
$\mathbb{S}_f^{\times \beta}$, with every subgroup $\mathbb{S}_f$ acting
on a particular sector $n_\mu$ of the decomposition of $n=\overline{n_1\dots n_\beta}$.
The general case of $N=\prod_{d=1}^D f_d^{\beta_d}$ (\refe{ham_gen_p1})
inherits its symmetries directly from the case $N=f^\beta$, simply producing
$\mathbb{S}_{f_1}^{\times\beta_1} \times\cdots\times \mathbb{S}_{f_D}^{\times\beta_D}$.

The symmetries for the non-primary phases $k, \gcd(N,k)\neq1$ (\refe{ham_gen_pa})
are the same as for the primary phases with $N'=N|k$
(the symmetry group is nominally smaller,
but it is important to note that the $\gcd(N,k)$ states of the original basis
were reduced to $\pi_x\in\{0,1\}$,
which defines an additional on-site symmetry $SU(\gcd(N,k)-1)$ in the original basis).

Apart from the already discussed global permutation symmetries,
the obtained systems also have symmetries with local charges.

\subsection{Winding and Laterality-like symmetries}

The prime-$N$ ($N=f$) Hamiltonian \refe{ham_prime} only allows transitions of
\begin{equation}
	\inb|{\cdots,n_0,n,n_0,\cdots}> \leftrightarrows \inb|{\cdots,n_0,n',n_0,\cdots}>	
\end{equation}
type. Then any ``symmetric charge" $q_{(x,y)}\defeq g_{()}(n_x,n_y)$
or any ``antisymmetric charge" $q_{[x,y]} \defeq g_{[]}(n_x,n_y)$
with an arbitrary symmetric function $g_{()}(n,m)=g_{()}(m,n)$
or an antisymmetric function $g_{[]}(n,m)=-g_{[]}(m,n)$, respectively,
gives rise to an integral of motion
\begin{equation}
	{\cal L} = \sum_{x\in\partial} (-1)^x q_{(x-1,x)} \squad,\squad
	{\cal W} = \sum_{x\in\partial} q_{[x-1,x]} \squad,
\end{equation}
as the change $n_x\rightarrow n_x'$ requires $n_{x-1}=n_{x+1}$,
which implies that $q_{{(x-1,x)}}-q_{{(x,x+1)}}=0$ and
$q_{{[x-1,x]}}+q_{{[x,x+1]}}=0$ both before and after the change.
The other charges are not affected.

There are $f(f+1)/2$ linearly independent symmetric functions $g_{()}$
and $f(f-1)/2$ antisymmetric functions $g_{[]}$.
We can use the basis $g_{(ab)}(n,m)=E_{ab}(n,m) + E_{ba}(n,m)$ and
$g_{[ab]}(n,m)=E_{ab}(n,m) - E_{ba}(n,m)$ where $a,b \in \mathbb{Z}_f$
with $E_{ab}(n,m)=\delta_{a,n} \delta_{b,m}$.
The functions of form $g(n)+g(m)$ and $g(n)-g(m)$ should be excluded
from the set of $g_{()}$-s and $g_{[]}$-s, respectively,
since they are ``exact" (and trivial due to the periodicity condition).
The corresponding basis sizes are $f$ and $f-1$.
The functions to be excluded from the set of basis elements $g_{(ab)}$
can be chosen to be the ones with $a=b$.
The number of the remaining independent motion integrals $\cal L$ and $\cal W$ is $(f-1)^2$.

One can notice that the global permutation symmetry group
${\cal S} = ({\cal S}_\text{e} \times {\cal S}_\text{o})_\text{diag}$ connects
all the $\cal L$-s and $\cal W$-s among themselves
as they map any generator $g_{(ab)}$ or $g_{[ab]}$ to the whole set.
The two base generators can be chosen as
\begin{equation}
\begin{split}
	g_{(0)}(n,m) &= \inb|{g_{[0]}(n,m)}| \squad, \\
	g_{[0]}(n,m) &= \sum_{k<f/2} [\delta_{n,m+k} - \delta_{n,m-k}] \cdot k/N
\end{split}
\end{equation}
where the addition/subtraction to $m$ is meant by $\text{mod } f$.
The bases $g_{(ab)}$ and $g_{[ab]}$ can be restored from $g_{(0)}$ and $g_{[0]}$
as linear combinations of $S^+ g_0 S$, $S \in \cal S$.
In the case $f=2$, there is no $\cal W$ (due to periodicity).
In the case $f=3$, the corresponding ${\cal W}_0$ and ${\cal L}_0$ are
the winding number and the laterality motion integrals
found in \cite{edge-z33}.

\begin{figure}
    \centering
    \includegraphics[width=.7\linewidth]{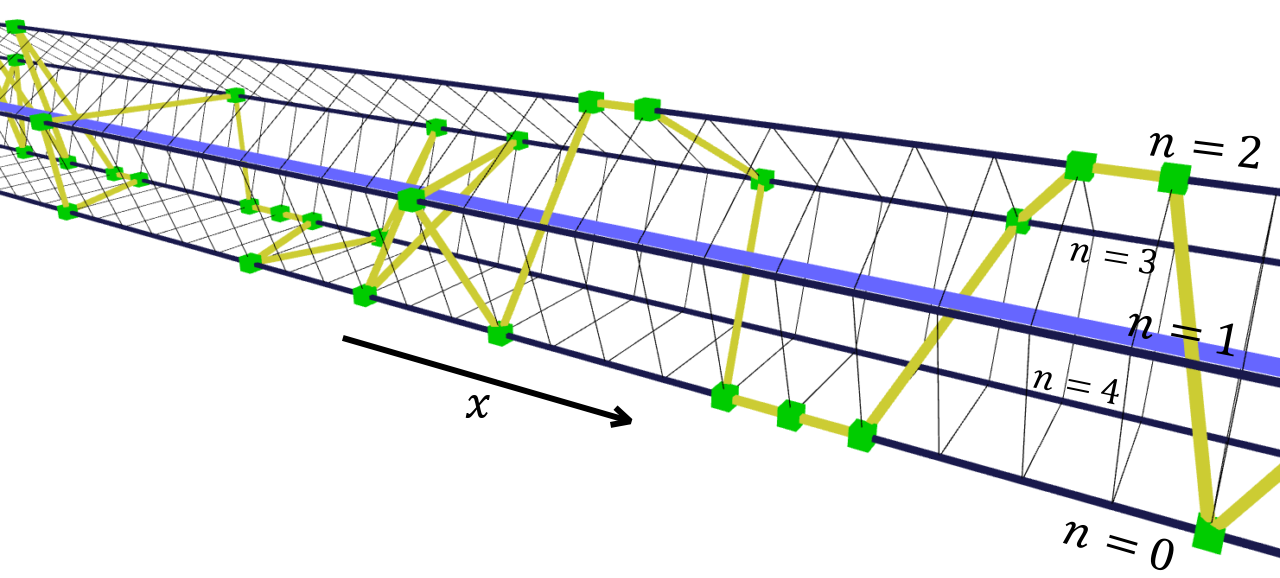}
    \caption{The visual representation of ${\cal W}_0$ for the case $f=5$.
        The axial lines representing different values of $n$ (thick dark lines)
        with the states $n_x$ at $x$ marked (green cubes) on them,
        and the polyline (yellow lines) connecting the markings.
        ${\cal W}_0$ is the number of revolutions of the polyline
        around the axis (light blue line).}
    \label{fig:winding}
    \vspace{-.5cm}
\end{figure}

${\cal W}_0$ can be given a visual interpretation (\reff{winding}).
Consider a cylinder with $f$ equally spaced axial lines around it,
enumerated from $0$ to $f-1$ consecutively
(non-consecutive enumerations will also produce integrals of motion,
just not the ${\cal W}_0$).
If we map the state $n_x$ for each $x$ to a point on the axial line with the number $n_x$
at position $x$ along the axis
and then connect the points corresponding to neighboring values of $x$,
the number of revolutions of the resulting polyline around the axis
will be the value of ${\cal W}_0$.

For a prime power $N=f^\beta$, the transitions allowed
by the primary phase Hamiltonian \refe{ham_prpow_p1} are
the changes $n_{\mu,x}\leftrightarrows n_{\mu,x}'$ for the states where
$n_{\mu',x-1} = n_{\mu',x+1}$ holds for all $\mu'\leq \mu$.
The integrals of motion that follow from this property are
\begin{equation}
	{\cal L} = \sum_{x\in\partial} (-1)^x q_{\mu(x-1,x)} \squad,\squad
	{\cal W} = \sum_{x\in\partial} q_{\mu[x-1,x]}
\end{equation}
where the charges are $q_{\mu(x,y)}=g_{()}(n_{\mu,x},n_{\mu,y})$ (symmetric) and
$q_{\mu[x,y]}=g_{[]}(n_{\mu,x},n_{\mu,y})$ (antisymmetric).
There are no nontrivial integrals of motion of similar form that feature both $\mu$ and $\mu'$.
Following the same logic as for the prime values of $N$,
the number of independent integrals of motion is $\beta (f-1)^2$.

Similarly to the globally charged symmetries,
the generic case $N=\prod_{d=1}^D f_d^{\beta_d}$ does not possess unique symmetries
but rather inherits them directly from its components with $N'=f_d^{\beta_d}$.

\subsection{Number of active elements}
\label{sec:num_act}

In this subsection, we will consider the prime values of $N=f$
and will work with the alternative representation defined by \refe{ham_prime_alt}.
For a given eigenstate of the basis $\bar{Z}_x$ given by the eigenvalues $\bar{n}_x$,
only the transitions of the form
\begin{equation}
	\inb|{\cdots,\bar{n},\bar{n},\cdots}> \leftrightarrows
    \inb|{\cdots,\bar{n}',\bar{n}',\cdots}>
    \label{eq:trans_alt}
\end{equation}
are allowed by the Hamiltonian.

The following representation of a state is introduced:
a value $c_x\in\{*,\op,\cl\}$ is assigned to each $x$;
$\bs{c}=[c_1,\dots,c_L]$ is a valid parentheses string (with periodic boundary conditions);
if a $\cl$ at position $x'$ matches the $\op$ at $x$
then $\bar{n}_x=\bar{n}_{x'}$ and there are no asterisks `$*$' between them;
the configuration is maximal, i.e.,
no asterisks can be replaced by parentheses while still conforming to the rules.
An example of this representation is shown below.
\begin{equation*}
\begin{split}
	\inb|{\bar{n}_1\dots\bar{n}_L}> = &\inb|{\text{\tt10101001220212201121}}> \rightarrow \\
    \rightarrow \bs{c} = &\inb[{\text{\tt)**((())())**()*()*(}}]
\end{split}
\end{equation*}

In general, the representation $\bs{c}$ of a state is not uniquely defined.
For instance, a sequence of three equal $\bar{n}$-s
can be marked as {\tt()*} and {\tt*()},
a sequence of four --- as {\tt()()} and {\tt(())}, etc.
More complicated examples include the sequence {\tt011001}
which can be depicted as {\tt*()()*}, {\tt(())**} or {\tt**(())}.
There are two important features of this depiction.
Firstly, if $\bar{n}_x=\bar{n}_{x+1}$,
there is a representation where $c_xc_{x+1}=\op\cl$
(as the construction of $\bs{c}$ can start with that pair).
Consequently, any two states with an allowed direct transition (\refe{trans_alt})
have a common depiction.
Secondly, any depiction of a given state has the same number of parentheses.
This combinatorial statement, although not proven mathematically,
is verified through numerical calculations.
Together with the first feature,
this implies that the number of parenthesis pairs $K$ is an integral of motion.

In a chain of transitions, nested parentheses are allowed to split and move:
a nested configuration {\tt(()(...))} corresponding to
$\bar n_1 \bar n_2 \bar  n_2 \bar n_3 \dots \bar n_k \bar n_1$
can transition into
$\bar n_1 \bar n_1 \bar  n_1 \bar n_3 \dots \bar n_k \bar n_1$,
depicted as {\tt()((...))}.
With a similar mechanism applied recursively,
the system can always reach a state without nesting.
In this case, each pair $\op\cl$ marks a pair of sites where the state is allowed to change,
thus making $K$ the number of independent morphable (active) elements.
These active elements can be interpreted as particles,
and the study of their statistics is a topic of further research.
The motion of a single particle (the sector $K=1$)
is discussed in detail in Appendix \ref{sec:dispers}.
It is important to note that the state with $K=0$ is not the ground state,
thus the ``particles" are not the low-energy excitations.

\subsection{The 't Hooft anomaly}

An important feature that sets apart the studied 1-dimensional chain
that was obtained as a boundary of a two-dimensional system from a standalone
1-dimensional chain is the presence of an anomalous symmetry inherited from the bulk.
The anomaly manifests itself as an obstruction to realizing the symmetry
as an ordinary local symmetry of the boundary theory alone,
indicating that it requires a higher-dimensional bulk for a consistent realization.
In the present case, we will study this anomaly directly from the boundary symmetry
by restricting the symmetry action to a finite or half-infinite portion of the boundary.

First, we need to obtain the expression of
the symmetry operator $S_{\bs{s}}$ on the boundary,
in the representation used for the boundary modes.
In this section, we will explicitly indicate the difference in representations
using the accented operator notation for the one used in the boundary modes
$\acute{\cal O}=U_k {\cal O} U_k^\dagger$.

In \cite{edge-z2f} the operators $V_{\bs{s}}^k$ are implicitly defined as
$S_{\bs{s}} U_k S_{\bs{s}}^\dagger U_k^\dagger$, which reduce to
\begin{equation}
    \inb.{V_{\bs{s}}^k}|_{\cal O} \equiv
    \prod_{\inb<{x,y}>\in \partial \cap\Omega_{\cal O}}
    \nu^{-k}(0, -\bs{s}, \bs{n}_x, \bs{n}_y)
\end{equation}
when acting on some operator $\acute{\cal O}$ as
$V_{\bs{s}}^k \acute{\cal O} V_{\bs{s}}^{-k}$.
Note that $\acute{n}_x^\alpha=n_x^\alpha$ and $\acute{V}_{\bs{s}}^k = V_{\bs{s}}^k$
since both commute with $U_k$.
Here, $\Omega_{\cal O}$ is the region consisting of the triangles that
contain the support of the operator $\cal O$
and $\partial\Omega_{\cal O}$ is the boundary of that region.
The product runs over the links that belong both to $\partial\Omega_{\cal O}$
and to the system boundary $\partial$.
The initial symmetry is then given by
\begin{equation}
    S_{\bs{s}} = S_{\bs{s}} U_k S_{\bs{s}}^\dagger U_k^\dagger U_k S_{\bs{s}} U_k^\dagger
    = V_{\bs{s}}^k \acute{S}_{\bs{s}}
\end{equation}
where $V_{\bs{s}}^k$ is the product on the whole boundary $\partial$.

The symmetry $S$, constructed in the accented operator representation,
is now dressed and is not a purely onsite symmetry.
Its anomaly can be explicitly seen when the symmetry is restricted
to an open finite section $\ell=[x_-,x_+]$ of the boundary
\cite{spt-cohom-thooft-finite, thooft-finite, edge-z3}.
In the bulk of $\ell$, the restricted symmetry contains the same local action
as the full boundary symmetry,
but additional terms localized near the two endpoints $\partial\ell$ are generated.
The anomaly is encoded in the nontrivial algebraic structure of the endpoint contributions.

The restricted symmetry operators $\check{S}_{\bs s}$ are given by
\begin{equation}
    \check{S}_{\bs s}=\prod_{\ell}
    \varepsilon^{k s^\A (n_x^\B n_{x+1}^\C - n_x^\C n_{x+1}^\B)}
    \cdot \prod_{\ell} \acute{X}_{x,\A}^{s^\A} \acute{X}_{x,\B}^{s^\B} \acute{X}_{x,\C}^{s^\C} \squad.
\end{equation}
These operators are a representation  of the initial symmetry group $S$ ($\bs{s}\in S$) 
with a nontrivial operator-valued multiplier.
Using the commutation relations of $\acute{X}_{x,\alpha}$ and $n_x^\alpha$,
their composition law can be derived to be
\begin{equation}
    \check{S}_{\bs a} \check{S}_{\bs b} =
    \Phi(\bs{a},\bs{b}) \cdot \check{S}_{\bs{a}+\bs{b}}
    \squad,\squad
    \Phi(\bs{a},\bs{b})=\Phi_-(\bs{a},\bs{b})\Phi_+(\bs{a},\bs{b})
    \squad,\squad \Phi_\pm(\bs{a},\bs{b})=
    \varepsilon^{\pm k b^\A (a^\B n_{x_\pm}^\C - a^\C n_{x_\pm}^\B)}
\end{equation}
where $\Phi_\pm(\bs{a},\bs{b})$ are localized at the two endpoints $x_\pm$.
The multipliers $\Phi_\pm(\bs{a},\bs{b})$ contribute nontrivial $U(1)$ phases to the associativity condition, demonstrating the anomaly. Generally,
\begin{equation}
   (\check{S}_{\bs a} \check{S}_{\bs b}) \check{S}_{\bs c} =
   \Phi(\bs{a},\bs{b},\bs{c}) \cdot
   \check{S}_{\bs a} (\check{S}_{\bs b} \check{S}_{\bs c}) \squad\Rightarrow\squad
    \Phi(\bs{a},\bs{b},\bs{c})=
    \frac{\Phi(\bs{a}+\bs{b},\bs{c}) \cdot \Phi(\bs{a},\bs{b})}
    {\check{S}_{\bs{a}} \Phi(\bs{b},\bs{c}) \check{S}_{\bs{a}}^{-1}
    \cdot \Phi(\bs{a},\bs{b}+\bs{c})} \squad.
\end{equation}
Here, the two endpoint contributions to the associator can be treated separately.
The phases $\Phi_\pm(\bs{a},\bs{b},\bs{c})$
produced by the corresponding multipliers $\Phi_\pm(\bs{a},\bs{b})$ are given by
\begin{equation}
    \Phi_\pm(\bs{a},\bs{b},\bs{c})=
    \varepsilon^{\pm kc^\A(a^\B b^\C  - b^\B a^\C)}
\end{equation}
which results in $\Phi(\bs{a},\bs{b},\bs{c})=1$.
As expected,
the overall associativity is retained despite nontrivial endpoint contributions.

The emergent nontrivial phase $\Phi_+(\bs{a},\bs{b}, \bs{c})$ is the cohomology element
$\nu^k(0,\bs{c},\bs{a},\bs{b})$ used for \refe{V2H}.
Similarly, $\Phi_-(\bs{a},\bs{b}, \bs{c})=\nu^k(0,\bs{c},\bs{b},\bs{a})$.
Hence, the restricted boundary symmetry directly realizes the 't Hooft anomaly,
represented by the corresponding nontrivial element of $H^3(S,U(1))$.

\section{Conclusion}

In this work, we have constructed and analyzed a family of one-dimensional systems
arising from the $\mathbb{Z}_N^{\times 3}$ symmetry protected topological phases in
a two-dimensional $\mathbb{Z}_N$ Potts model on a triangular lattice.
Using a cohomology-based transformation, we obtained explicit boundary Hamiltonians
describing the boundary modes and demonstrated that their structure strongly depends
on the arithmetic properties of $N$.

We showed that for prime values of $N$ the boundary theory simplifies to a
symmetry-rich constrained system and admits a formulation in terms of
mutually commuting Temperley–Lieb algebras.
For prime power and general composite values of $N$,
the models exhibit a hierarchical and factorized structure, respectively.

An important result of this work is that the nontrivial boundary theories in this class
can be reduced to primary models supplemented by local defect degrees of freedom.
These defects act as dynamical constraints
that partition the system into independent segments.
This provides a unified description of all phases
except the odd-numbered phases for even values of $N$.

We further analyzed the symmetry properties of the boundary theories
and demonstrated the presence of an anomalous implementation of
the global $\mathbb{Z}_N^{\times 3}$ symmetry.
We obtained a representation with a nontrivial associator
that directly reproduces the underlying group cohomology 3-cocycle,
thus providing a concrete lattice realization of the corresponding ’t Hooft anomaly.

The explicit structure of the boundary Hamiltonians and their connection to
Temperley–Lieb algebras suggest possible links to integrable systems and
conformal field theories describing the continuum limit.
Exploring these connections and investigating the possible
particle-like excitations associated with the symmetry currents and their statistics
remains an interesting direction for future work.

\appendix

\section{Integrable extension of $\mathbb{Z}_f^{\times 3}$ SPT boundary modes}
\label{sec:integr}

An integrable extension of the model given by \refe{ham_TL} can be constructed
using the $R$-matrix formalism, using the $R$-matrix ansatz
\begin{equation}
    R(u) = a_u\ \raisebox{-1.4ex}{\includegraphics[height=4ex]{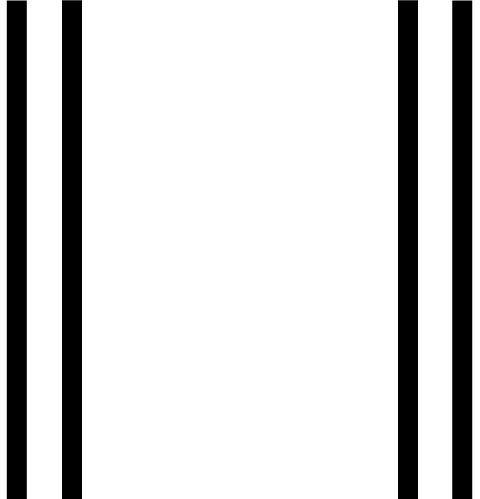}}
        + b_u\ \raisebox{-1.4ex}{\includegraphics[height=4ex]{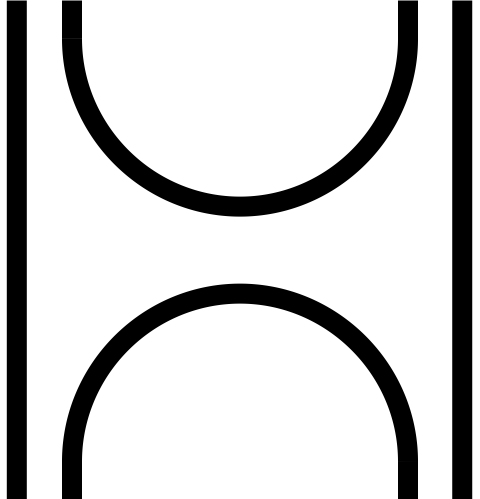}}
        + c_u\ \raisebox{-1.4ex}{\includegraphics[height=4ex]{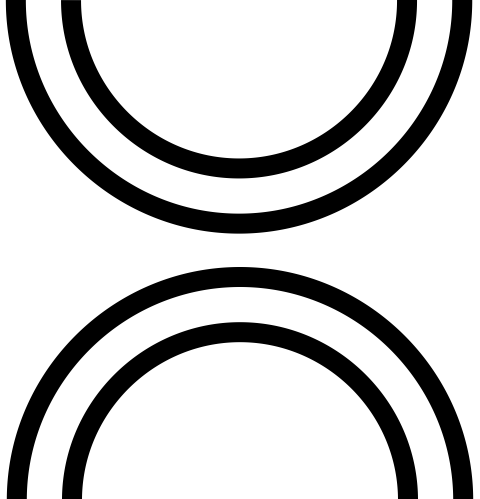}} \squad.
    \label{eq:Rform}
\end{equation}
The diagrams are presented in the regular Temperley-Lieb formalism,
i.e. the product of the operators is given by stacking diagrams,
diagrams are defined up to continuous deformations of lines,
and all closed loops are equivalent to scalar multiplication by the loop weight.
In our case, the loop weight is $\lambda=\sqrt{f}$.
The Yang-Baxter equation (YBE) is given by
\begin{equation}
    R_{23}(v) \hat{R}_{12}(u,v) R_{23}(u) = R_{12}(u) \hat{R}_{23}(u,v) R_{12}(v)
    \label{eq:YBE_init}
\end{equation}
where $\hat{R}(u,v)$ is the intertwining matrix of a form similar to \refe{Rform},
with replaced $(a,b,c)_u \rightarrow (\hat{a},\hat{b},\hat{c})_{u,v}$.
The lower indices of $R$ and $\hat{R}$ indicate the sites of their action.

For simplicity, a notation system for all 2- and 3-site diagrams is introduced as follows:
\begin{table}[h]
\centering
\setlength{\tabcolsep}{29.9pt}
\begin{tabular}{c|c|c}
    $\I$ & $\h$ & $\X$
    \\\hline\vspace{-8pt}&&\\ 
    $\raisebox{-1.4ex}{\includegraphics[height=4ex]{diags/I}}$ &
    $\raisebox{-1.4ex}{\includegraphics[height=4ex]{diags/H}}$ &
    $\raisebox{-1.4ex}{\includegraphics[height=4ex]{diags/X}}$
    \\\vspace{-8pt}&&\\
\end{tabular}
\setlength{\tabcolsep}{5.6pt}
\begin{tabular}{c|c|c|c|c|c}
    \hline$\alpha$ & $\beta_1$ & $\beta_2$ & $\gamma_1$ & $\gamma_2$ & $\delta$
    \\\hline\vspace{-8pt}&&&&&\\ 
    $\raisebox{-1.4ex}{\includegraphics[height=4ex]{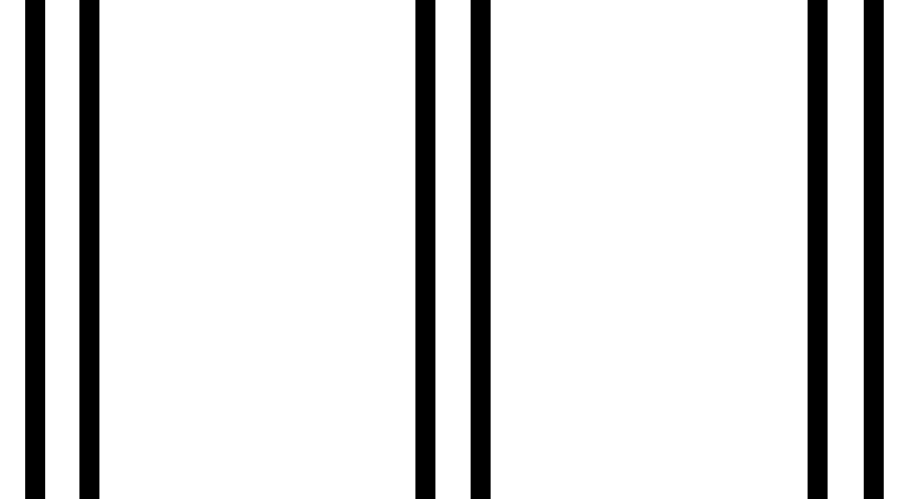}}$ &
    $\raisebox{-1.4ex}{\includegraphics[height=4ex]{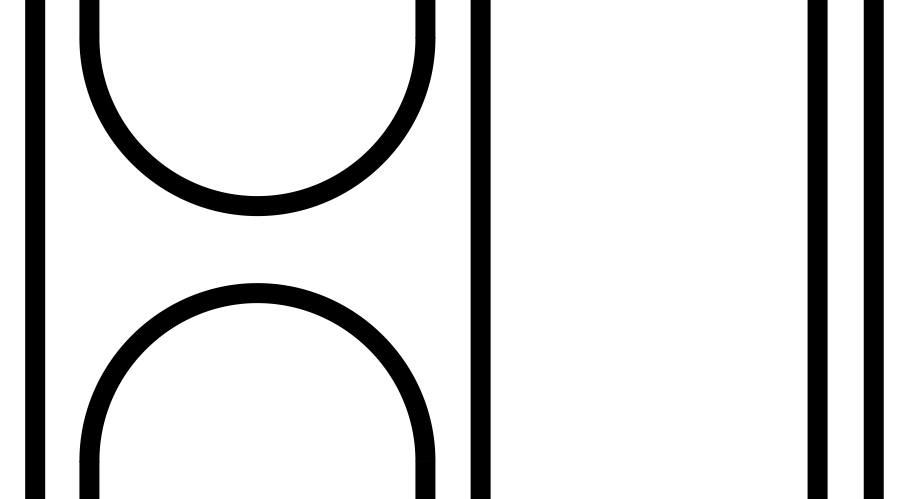}}$ &
    $\raisebox{-1.4ex}{\reflectbox{\includegraphics[height=4ex]{diags/b}}}$ &
    $\raisebox{-1.4ex}{\includegraphics[height=4ex]{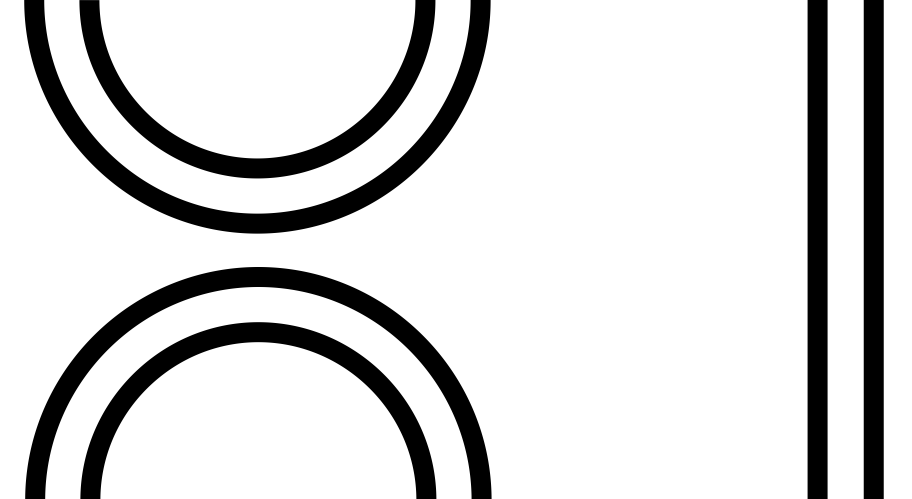}}$ &
    $\raisebox{-1.4ex}{\reflectbox{\includegraphics[height=4ex]{diags/g}}}$ &
    $\raisebox{-1.4ex}{\includegraphics[height=4ex]{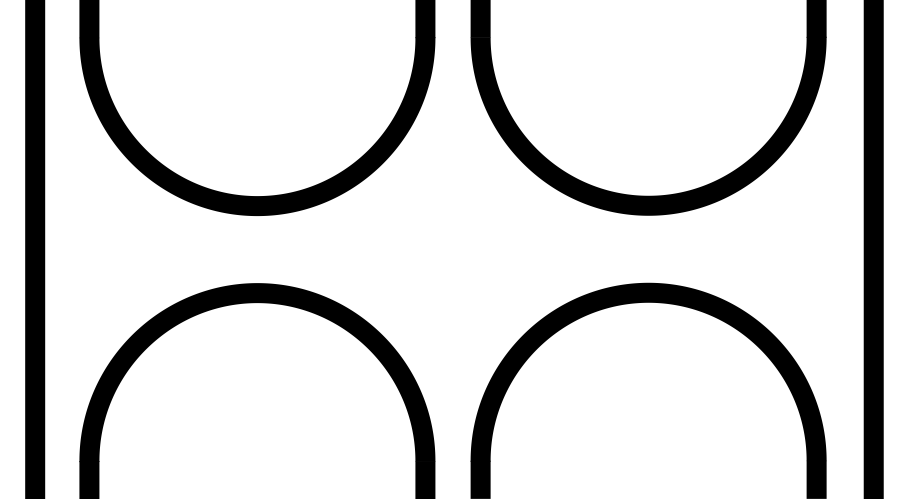}}$
    \\\vspace{-8pt}&&&&&\\\hline
    $\varepsilon_1$ & $\varepsilon_2$ & $\varepsilon_3$ & $\varepsilon_4$ &
    $\zeta_1$ & $\zeta_2$
    \\\hline\vspace{-8pt}&&&&&\\
    $\raisebox{-1.4ex}{\includegraphics[height=4ex]{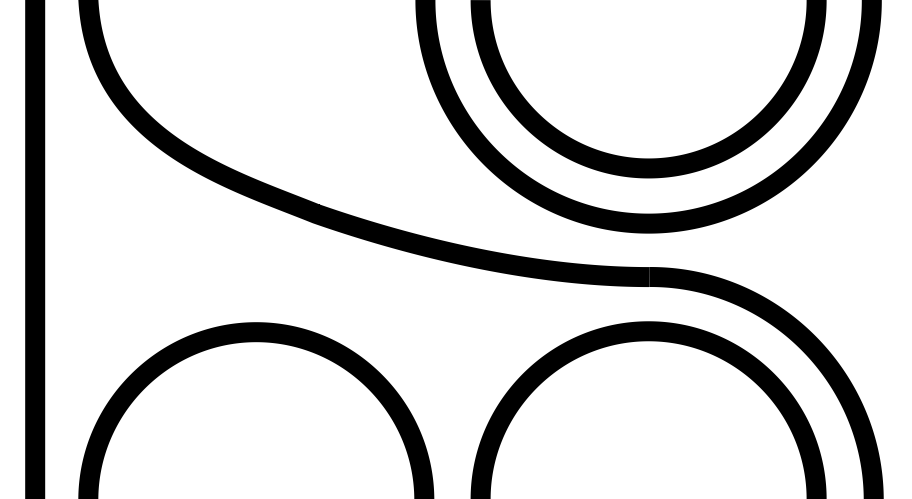}}$ &
    $\raisebox{-1.4ex}{\scalebox{-1}[1]{\includegraphics[height=4ex]{diags/e}}}$ &
    $\raisebox{2.6ex}{\scalebox{1}[-1]{\includegraphics[height=4ex]{diags/e}}}$ &
    $\raisebox{2.6ex}{\scalebox{-1}[-1]{\includegraphics[height=4ex]{diags/e}}}$ &
    $\raisebox{-1.4ex}{\includegraphics[height=4ex]{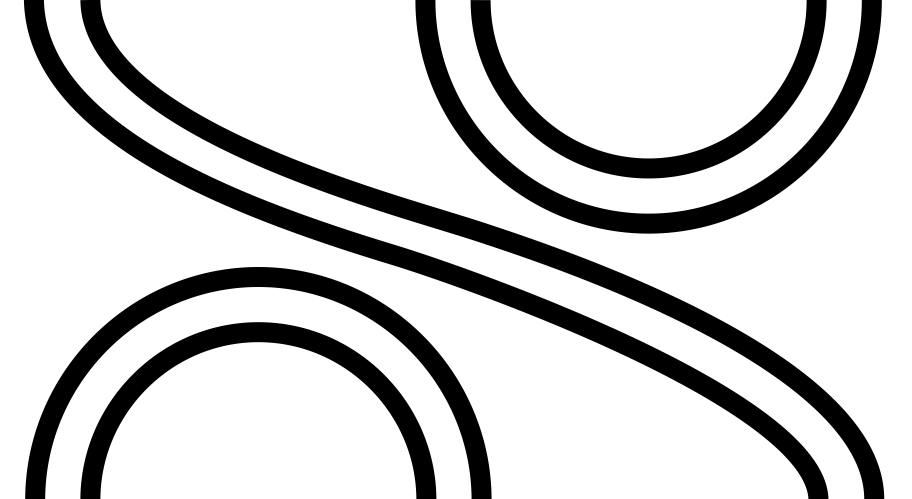}}$ &
    $\raisebox{-1.4ex}{\reflectbox{\includegraphics[height=4ex]{diags/z}}}$ 
\end{tabular}
\end{table}

It is also useful to define a ``multiplication table" for the expressions
of form $C_{23}B_{12}A_{23}$ producing 3-site diagrams:
\begin{table}[h]
\centering
\setlength{\tabcolsep}{4.5pt}
\begin{tabular}{c|ccc|ccc|ccc}
     $B$ & \multicolumn{3}{c|}{$\I$}
         & \multicolumn{3}{c|}{$\h$} & \multicolumn{3}{c}{$\X$} \\ \hline
    $A \backslash C$ &$\I$&$\h$&$\X$ &$\I$&$\h$&$\X$ &$\I$&$\h$&$\X$ \\ \hline
    $\I$ & $\alpha$ & $\beta_2$ & $\gamma_2$
        & $\beta_1$ & $\delta$ & $\varepsilon_1$
        & $\gamma_1$ & $\varepsilon_4$ & $\zeta_1$ \\
    $\h$ & $\beta_2$ & $\lambda\beta_2$ & $\lambda\gamma_2 $
        & $\delta$ & $\lambda\delta$ & $\lambda\varepsilon_1$ 
        & $\varepsilon_2$ & $\delta$ & $\varepsilon_1$  \\
    $\X$ & $\gamma_2$ & $\lambda\gamma_2$ & $\lambda^2\gamma_2$
        & $\varepsilon_3$ & $\lambda\varepsilon_3$ & $\lambda\gamma_2$
        & $\zeta_2$ & $\varepsilon_3$ & $\gamma_2$
\end{tabular}
\end{table}

\pagebreak

\refe{YBE_init} then takes the form
\begin{equation}
    \sum_\kappa g_l^\kappa(u,v) \kappa = \sum_\kappa g_r^\kappa(u,v) \kappa
\end{equation}
where $\kappa$ enumerates all possible diagrams, $\kappa\in\{\alpha,\beta_1,\dots,\zeta_2\}$,
and $g_{l/r}^\kappa(u,v)$ are the corresponding coefficients on the
left and right side of \refe{YBE_init}, which can be calculated directly.
The equation is then reduced to $g_l^\kappa(u,v)=g_r^\kappa(u,v)$

Then we define the parity (horizontal reflection) operator $P$ and
the time reversal (vertical reflection) operator $T$ for the diagrams,
and introduce the notation $\bar\kappa=PT\kappa$.
In general, there is a term $g(u,v) PT\inb({\bar A_{12} \bar B_{23} \bar C_{12}})$
on the right side of \refe{YBE_init} corresponding to each left side term
$g(u,v) C_{23}B_{12}A_{23}$.
Since in our case $\bar\I=\I$, $\bar\h=\h$ and $\bar\X=\X$,
then $g_l^\kappa(u,v)=g_r^{\bar\kappa}(u,v)$, and therefore the YBE becomes
\begin{equation}
    g_l^\kappa(u,v) = g_l^{\bar\kappa}(u,v)
\end{equation}
for all $\kappa$.
Some of those equations are trivial since
$\bar\alpha=\alpha$, $\bar\delta=\delta$ and $\bar\zeta_{1/2}=\zeta_{1/2}$.
The non-trivial ones come from $PT$ mappings
$\beta_1\leftrightarrow\beta_2$, $\gamma_1\leftrightarrow\gamma_2$,
$\varepsilon_1\leftrightarrow\varepsilon_4$ and $\varepsilon_2\leftrightarrow\varepsilon_3$,
and are
\begin{equation}
\begin{split}
    \beta:\ (&a_u b_v + b_u a_v + \lambda b_u b_v) \hat a_{u,v} - a_u a_v \hat b_{u,v} = 0
    \squad,\\
    \gamma:\ (&a_u c_v + c_u a_v + \lambda b_u c_v + \lambda c_u b_v) \hat a_{u,v}\\
        +&c_u c_v (\lambda^2 \hat a_{u,v} + \lambda \hat b_{u,v} + \hat c_{u,v})
        - a_u a_v \hat c_{u,v}=0 \squad,\\
    \varepsilon:\ (&a_u + \lambda b_u) c_v \hat b_{u,v} + (b_u c_v - a_u b_v) \hat c_{u,v} = 0
\end{split}
\end{equation}
(the two equations coming from $\varepsilon_1\leftrightarrow\varepsilon_4$ and
$\varepsilon_2\leftrightarrow\varepsilon_3$ are equivalent).
The equations can be written in a matrix form
\begin{equation}
    \inb({\begin{matrix}
        b_v + b_u + \lambda b_u b_v & -1 & 0 \\
        c_v+c_u+\lambda(b_uc_v+c_ub_v)+\lambda^2c_uc_v & \lambda c_u c_v & c_u c_v - 1\\
        0 & c_v (1 + \lambda b_u) & b_u c_v - b_v
    \end{matrix}})
    \inb({\begin{matrix}
        \hat a_{u,v} \\ \hat b_{u,v} \\ \hat c_{u,v}
    \end{matrix}}) = 0
\end{equation}
where we used the freedom of normalization of $a_u$ to unity
without loss of generality.

The equation will have a nontrivial solution for $(\hat a_{u,v}, \hat b_{u,v}, \hat c_{u,v})$
iff the determinant $\det_{u,v}$ of the coefficient matrix is $0$.
Using the small $v$ expansion $b(v)=b^{(0)}+b^{(1)}v +\mathcal{O}(v^2)$ and
$c(v)=c^{(0)}+c^{(1)}v +\mathcal{O}(v^2)$,
one can solve the order-$0$ equation and find a simple particular solution
\begin{equation}
    {\det}_{u,0} = 0 \squad \Leftarrow \squad b^{(0)}=c^{(0)}=0 \squad.
\end{equation}
Under this constraint, the order-$1$ equation provides
\begin{equation}
    \partial_v{\det}_{u,v} \Big|_{v=0} = 0 \squad \Leftrightarrow \squad
    c_u= b_u \frac{1+\lambda b_u}{\rho-b_u}\equiv\varphi(b_u)
\end{equation}
where $\rho=b^{(1)}/c^{(1)}$.
The result is consistent with the expansion.
Finally, expressing $c_u$ and $c_v$ by $b_u$ and $b_v$ via the obtained relation
factorizes the general expression of $\det_{u,v}$,
\begin{equation}
    {\det}_{u,v} \Big|_{c=\varphi(b)} = \frac {b_u b_v^2 (b_u+b_v+\lambda b_u b_v-\rho)
    (\lambda^2-\lambda\rho-2) (1+\lambda b_u) (1+\lambda b_v)}{(b_u-\rho)(b_v-\rho)^2}=0\squad.
\end{equation}
Notice that the above relation does not contain approximations.
Apart from the trivial solutions as $b_u=0$
(which is equivalent to the YBE solution with a single-line TL algebra),
the equation is satisfied for
\begin{equation}
    \rho=(\lambda^2-2)/\lambda \squad.
    \label{eq:integr}
\end{equation}

This $R$-matrix produces Hamiltonians of form
\begin{equation}
    H_I=\sum_x \partial_u R_x(u) \Big|_{u=0} \propto \sum_x
    \bigg[\raisebox{-1.4ex}{\includegraphics[height=4ex]{diags/X}}
        + \rho\ \raisebox{-1.4ex}{\includegraphics[height=4ex]{diags/H}} \bigg]_x
    \label{eq:ham_R}
\end{equation}
which are integrable if $\rho$ is given by \refe{integr}.

The SPT Hamiltonian \refe{ham_TL} is reproduced with $\rho=-1/\sqrt{f}$,
which satisfies the integrability condition only if $f=1$, which is a trivial result.

\section{Dispersion relation of a single emergent quasiparticle
 in $\mathbb{Z}_f^{\times 3}$ SPT boundary modes}
\label{sec:dispers}

Here, we consider a single quasiparticle introduced in Sec.\ref{sec:num_act}
and derive its dispersion relation.
Switching to the initial Hamiltonian formalism given by \refe{ham_prime},
the states of the sector $K=1$ are described by the condition
$n_{x-1} = n_{x+1} \Leftrightarrow \bar{n}_{x-1}=\bar{n}_x$ for a single value of $x$.
Additionally, $n_x = n_{x\pm 2}$ is allowed
(it corresponds to the case of three consequent equal $\bar n$-s), and
$n_{x-2} \neq n_{x+2}$ is required
(otherwise the state will correspond to a nested configuration with $K=2$).

The states $K=1$ will be represented as $\inb|{\bs\alpha; x, \beta}>$,
where $\bs\alpha$ is an array of $\mathbb{Z}_f$ elements
of size $L'=L-2$ (enumerated from $1$ to $L'$)
and $\beta$ is another $\mathbb{Z}_f$ element:
\begin{equation}
    \inb|{\bs\alpha; x, \beta}> \defeq
    \inb|{\alpha_1, \cdots, \alpha_x, \beta, \alpha_x, \cdots, \alpha_{L'}}> \squad.
\end{equation}
The constraints stated above are imposed as
$\alpha_{x'-1} \neq \alpha_{x'+1}$ for all $x'$.
However, in the case $\beta=\alpha_{x\pm 1}$ the notation is no longer unique since
\begin{equation}
    \inb|{\bs\alpha; x, \alpha_{x+1}}>=\inb|{\bs\alpha; x+1, \alpha_x}> \defeq \inb|{\bs\alpha;x+\half}> \squad.
\end{equation}
Thus, the basis dimension is $f-1$ per $x$.

The Hamiltonian will act nontrivially on the $K=1$ states:
for the $\beta\neq\alpha_{x\pm 1}$ and $\beta=\alpha_{x\pm 1}$ type states
the corresponding actions are
\begin{equation}
\begin{split}
    &H_\partial^{f} \inb|{\bs\alpha; x, \beta}> \xrightarrow{\beta\neq\alpha_{x\pm 1}}
    -\widetilde{\mathfrak{h}}_x^f \inb|{\bs\alpha; x, \beta}> = 
    -\sum_{\beta'\neq\beta} \inb|{\bs\alpha; x, \beta'}> \squad,\\
    &H_\partial^f \inb|{\bs\alpha;x+\half}> =
    -(\widetilde{\mathfrak{h}}_x^f+\widetilde{\mathfrak{h}}_{x+1}^f)
    \inb|{\bs\alpha;x+\half}> = -\sum_{\beta'}
    \Big( \inb|{\bs\alpha;x,\beta'}> + \inb|{\bs\alpha;x+1,\beta'}> \Big) +
    2 \inb|{\bs\alpha;x+\half}> \squad .
\end{split}
\end{equation}
and thus the space of $\inb|{\bs\alpha;x,\beta}>$ for a given $\bs\alpha$
is closed under $H_\partial^f$.

As a result of different space dimensions of $n$ and $\bs\alpha$
($n_{x+L}=n_x$ but $\alpha_{x+L}=\alpha_{x+2}$)
the periodicity condition is not trivial here.
It is manifested as the shifting of the background $\bs\alpha$
when the ``particle" completes a turn around the chain.
The periodicity condition is therefore
$\inb|{\bs\alpha;x+L,\beta}>=\inb|{\bs\alpha^{(2)};x,\beta}>$,
where $\bs\alpha^{(d)}$ is the $\bs\alpha$ shifted left by $d$,
i.e. $\bs\alpha^{(d)}_{x-d}=\bs\alpha_x$.
Thus, the states are generally periodic in $x$ with period $LL'/2$.
Additional symmetries in $\bs\alpha$ can decrease the periodicity.

One can check that the states of form
$\inb|{\bs\alpha;x,\beta_2}>-\inb|{\bs\alpha;x,\beta_1}>$
with $\beta_{1/2}\neq\alpha_{x\pm1}$ are eigenstates with energy $1$.
There are $f-3$ such independent states per $x$.
The remaining $2$ come as delocalized states.

We can define $\inb|{\bs\alpha;x}>=\sum_{\beta'} \inb|{\bs\alpha;x,\beta'}>$.
With these notations, the Hamiltonian action will be given as
\begin{equation}
\begin{split}
    H_\partial^f \inb|{\bs\alpha;x}>&=-(f-1)\inb|{\bs\alpha;x}>
        -\inb|{\bs\alpha;x-1}>-\inb|{\bs\alpha;x+1}>
        +\inb|{\bs\alpha;x-\half}>+\inb|{\bs\alpha;x+\half}> \squad,\\
    H_\partial^f \inb|{\bs\alpha;x+\half}>&=-\inb|{\bs\alpha;x}>-\inb|{\bs\alpha;x+1}>
        +2\inb|{\bs\alpha;x+\half}> \squad,
\end{split}
\end{equation}
which essentially constitutes a simple hopping model
for which the eigenstates can be found in the form
\begin{equation}
\begin{split}
    \inb|{\bs\alpha;p}> \defeq \sum_x e^{i px} \Big(
        &\inb|{\bs\alpha;x}> + e^{i p/2} A \inb|{\bs\alpha;x+\half}> \Big) \squad,\\
    H \inb|{\bs\alpha;p}> = \sum_x e^{i px} \Big(
        &\inb[{-f+1-2\cos{p}-2A \cos{\frac{p}{2}}}] \inb|{\bs\alpha;x}> +
        e^{i p/2} A \inb[{2A^{-1}\cos{\frac{p}{2}} +2}] \inb|{\bs\alpha;x+\half}> \Big)
\end{split}
\end{equation}
and the eigenstate condition (the equality of the two square braces)
gives the values of $A$ and thus the energies, which are
\begin{equation}
    E_{p,\pm}=2\inb({\cos{\frac{p}{2}}\Big/ A_{p,\pm}+1}) \squad,\squad
    A_{p,\pm}=-b_p\pm\sqrt{b_p^2-1} \squad,\squad
    b_p=\cos{\frac{p}{2}} + \frac{f-1}{4} \Big/ \cos{\frac{p}{2}} \squad.
\end{equation}
The allowed values of $p$ are determined by the periodicity condition,
which yields $p\in \{4\pi/LL' \cdot \mathbb{N}\}\cap(-\pi,\pi]$.
For $f\to\infty$, $E_{p,-}=2$, $E_{p,+}=-f+1-2\cos{p}$.

\ack{I am grateful to
Tigran Hakobyan, Vadim Ohanyan, Hrachya Babujian, Ara Sedrakyan and Tigran Sedrakyan
for the valuable discussions.}

\funding{The research was supported by Armenian HECS Remote Laboratory program 24RL-1C024.}

\data{No new data were created or analyzed in this study.}

\bibliographystyle{iopart-num}
\bibliography{refs}

@article{spt-orig-1,
  title = {Local unitary transformation,  long-range quantum entanglement,  wave function renormalization,  and topological order},
  volume = {82},
  ISSN = {1550-235X},
  url = {http://dx.doi.org/10.1103/PhysRevB.82.155138},
  DOI = {10.1103/physrevb.82.155138},
  number = {15},
  journal = {Physical Review B},
  publisher = {American Physical Society (APS)},
  author = {Chen,  Xie and Gu,  Zheng-Cheng and Wen,  Xiao-Gang},
  year = {2010},
  month = oct 
}

@article{spt-orig-2,
  title = {Symmetry-Protected Topological Orders in Interacting Bosonic Systems},
  volume = {338},
  ISSN = {1095-9203},
  url = {http://dx.doi.org/10.1126/science.1227224},
  DOI = {10.1126/science.1227224},
  number = {6114},
  journal = {Science},
  publisher = {American Association for the Advancement of Science (AAAS)},
  author = {Chen,  Xie and Gu,  Zheng-Cheng and Liu,  Zheng-Xin and Wen,  Xiao-Gang},
  year = {2012},
  month = dec,
  pages = {1604–1606}
}

@article{spt-orig-3,
  title = {Nonlinear Field Theory of Large-Spin Heisenberg Antiferromagnets: Semiclassically Quantized Solitons of the One-Dimensional Easy-Axis N\'eel State},
  author = {Haldane, F. D. M.},
  journal = {Phys. Rev. Lett.},
  volume = {50},
  issue = {15},
  pages = {1153--1156},
  numpages = {0},
  year = {1983},
  month = {Apr},
  publisher = {American Physical Society},
  doi = {10.1103/PhysRevLett.50.1153},
  url = {https://link.aps.org/doi/10.1103/PhysRevLett.50.1153}
}

@article{spt-rev-1,
  title = {Symmetry-Protected Topological Phases of Quantum Matter},
  volume = {6},
  ISSN = {1947-5462},
  url = {http://dx.doi.org/10.1146/annurev-conmatphys-031214-014740},
  DOI = {10.1146/annurev-conmatphys-031214-014740},
  number = {1},
  journal = {Annual Review of Condensed Matter Physics},
  publisher = {Annual Reviews},
  author = {Senthil,  T.},
  year = {2015},
  month = mar,
  pages = {299–324}
}

@article{topol-1,
  title = {Detecting Topological Order in a Ground State Wave Function},
  author = {Levin, Michael and Wen, Xiao-Gang},
  journal = {Phys. Rev. Lett.},
  volume = {96},
  issue = {11},
  pages = {110405},
  numpages = {4},
  year = {2006},
  month = {Mar},
  publisher = {American Physical Society},
  doi = {10.1103/PhysRevLett.96.110405},
  url = {https://link.aps.org/doi/10.1103/PhysRevLett.96.110405}
}

@article{topol-2,
  title = {Topological Entanglement Entropy},
  author = {Kitaev, Alexei and Preskill, John},
  journal = {Phys. Rev. Lett.},
  volume = {96},
  issue = {11},
  pages = {110404},
  numpages = {4},
  year = {2006},
  month = {Mar},
  publisher = {American Physical Society},
  doi = {10.1103/PhysRevLett.96.110404},
  url = {https://link.aps.org/doi/10.1103/PhysRevLett.96.110404}
}

@article{edge-levin_gu,
  title = {Braiding statistics approach to symmetry-protected topological phases},
  volume = {86},
  ISSN = {1550-235X},
  url = {http://dx.doi.org/10.1103/PhysRevB.86.115109},
  DOI = {10.1103/physrevb.86.115109},
  number = {11},
  journal = {Physical Review B},
  publisher = {American Physical Society (APS)},
  author = {Levin,  Michael and Gu,  Zheng-Cheng},
  year = {2012},
  month = sep 
}

@article{pract-1,
  title = {Symmetry-protected topologically ordered states for universal quantum computation},
  author = {Nautrup, Hendrik Poulsen and Wei, Tzu-Chieh},
  journal = {Phys. Rev. A},
  volume = {92},
  issue = {5},
  pages = {052309},
  numpages = {17},
  year = {2015},
  month = {Nov},
  publisher = {American Physical Society},
  doi = {10.1103/PhysRevA.92.052309},
  url = {https://link.aps.org/doi/10.1103/PhysRevA.92.052309}
}

@article{pract-2,
  title = {Computationally Universal Phase of Quantum Matter},
  author = {Raussendorf, Robert and Okay, Cihan and Wang, Dong-Sheng and Stephen, David T. and Nautrup, Hendrik Poulsen},
  journal = {Phys. Rev. Lett.},
  volume = {122},
  issue = {9},
  pages = {090501},
  numpages = {5},
  year = {2019},
  month = {Mar},
  publisher = {American Physical Society},
  doi = {10.1103/PhysRevLett.122.090501},
  url = {https://link.aps.org/doi/10.1103/PhysRevLett.122.090501}
}

@article{pract-3,
  title = {Edge Modes and Symmetry-Protected Topological States in Open Quantum Systems},
  author = {Paszko, Dawid and Rose, Dominic C. and Szyma\ifmmode \acute{n}\else \'{n}\fi{}ska, Marzena H. and Pal, Arijeet},
  journal = {PRX Quantum},
  volume = {5},
  issue = {3},
  pages = {030304},
  numpages = {22},
  year = {2024},
  month = {Jul},
  publisher = {American Physical Society},
  doi = {10.1103/PRXQuantum.5.030304},
  url = {https://link.aps.org/doi/10.1103/PRXQuantum.5.030304}
}

@article{pract-z23,
  title = {Hierarchy of universal entanglement in 2D measurement-based quantum computation},
  volume = {2},
  ISSN = {2056-6387},
  url = {http://dx.doi.org/10.1038/npjqi.2016.36},
  DOI = {10.1038/npjqi.2016.36},
  number = {1},
  journal = {npj Quantum Information},
  publisher = {Springer Science and Business Media LLC},
  author = {Miller,  Jacob and Miyake,  Akimasa},
  year = {2016},
  month = nov 
}

@article{pract-zn3,
  title = {Universal quantum computing using ${({\mathbb{Z}}_{d})}^{3}$ symmetry-protected topologically ordered states},
  author = {Chen, Yanzhu and Prakash, Abhishodh and Wei, Tzu-Chieh},
  journal = {Phys. Rev. A},
  volume = {97},
  issue = {2},
  pages = {022305},
  numpages = {20},
  year = {2018},
  month = {Feb},
  publisher = {American Physical Society},
  doi = {10.1103/PhysRevA.97.022305},
  url = {https://link.aps.org/doi/10.1103/PhysRevA.97.022305}
}

@Article{spt-cohom-1,
author={Kapustin, Anton
and Turzillo, Alex},
title={Equivariant topological quantum field theory and symmetry protected topological phases},
journal={Journal of High Energy Physics},
year={2017},
month={Mar},
day={01},
volume={2017},
number={3},
pages={6},
issn={1029-8479},
doi={10.1007/JHEP03(2017)006},
url={https://doi.org/10.1007/JHEP03(2017)006}
}

@article{spt-cohom-2,
  title = {Symmetry-protected topological orders for interacting fermions: Fermionic topological nonlinear $\ensuremath{\sigma}$ models and a special group supercohomology theory},
  author = {Gu, Zheng-Cheng and Wen, Xiao-Gang},
  journal = {Phys. Rev. B},
  volume = {90},
  issue = {11},
  pages = {115141},
  numpages = {59},
  year = {2014},
  month = {Sep},
  publisher = {American Physical Society},
  doi = {10.1103/PhysRevB.90.115141},
  url = {https://link.aps.org/doi/10.1103/PhysRevB.90.115141}
}

@article{spt-cohom-3,
  title = {Symmetry protected topological orders and the group cohomology of their symmetry group},
  author = {Chen, Xie and Gu, Zheng-Cheng and Liu, Zheng-Xin and Wen, Xiao-Gang},
  journal = {Phys. Rev. B},
  volume = {87},
  issue = {15},
  pages = {155114},
  numpages = {48},
  year = {2013},
  month = {Apr},
  publisher = {American Physical Society},
  doi = {10.1103/PhysRevB.87.155114},
  url = {https://link.aps.org/doi/10.1103/PhysRevB.87.155114}
}

@article{spt-cohom-thooft-finite,
  title = {Classifying symmetry-protected topological phases through the anomalous action of the symmetry on the edge},
  author = {Else, Dominic V. and Nayak, Chetan},
  journal = {Phys. Rev. B},
  volume = {90},
  issue = {23},
  pages = {235137},
  numpages = {19},
  year = {2014},
  month = {Dec},
  publisher = {American Physical Society},
  doi = {10.1103/PhysRevB.90.235137},
  url = {https://link.aps.org/doi/10.1103/PhysRevB.90.235137}
}

@article{spt-beyond-1,
  title = {Colloquium: Zoo of quantum-topological phases of matter},
  author = {Wen, Xiao-Gang},
  journal = {Rev. Mod. Phys.},
  volume = {89},
  issue = {4},
  pages = {041004},
  numpages = {17},
  year = {2017},
  month = {Dec},
  publisher = {American Physical Society},
  doi = {10.1103/RevModPhys.89.041004},
  url = {https://link.aps.org/doi/10.1103/RevModPhys.89.041004}
}

@Article{spt-beyond-2,
author={Inamura, Kansei},
title={Topological field theories and symmetry protected topological phases with fusion category symmetries},
journal={Journal of High Energy Physics},
year={2021},
month={May},
day={21},
volume={2021},
number={5},
pages={204},
issn={1029-8479},
doi={10.1007/JHEP05(2021)204},
url={https://doi.org/10.1007/JHEP05(2021)204}
}

@article{spt-beyond-3,
  title = {Gapped boundary of $(4+1)\mathrm{d}$ beyond-cohomology bosonic SPT phase},
  author = {Yang, Xinping and Cheng, Meng},
  journal = {Phys. Rev. B},
  volume = {110},
  issue = {4},
  pages = {045137},
  numpages = {16},
  year = {2024},
  month = {Jul},
  publisher = {American Physical Society},
  doi = {10.1103/PhysRevB.110.045137},
  url = {https://link.aps.org/doi/10.1103/PhysRevB.110.045137}
}

@article{edge-z33,
  title = {Z3 and (×Z3)3 symmetry protected topological paramagnets},
  volume = {2023},
  ISSN = {1029-8479},
  url = {http://dx.doi.org/10.1007/JHEP12(2023)199},
  DOI = {10.1007/jhep12(2023)199},
  number = {12},
  journal = {Journal of High Energy Physics},
  publisher = {Springer Science and Business Media LLC},
  author = {Topchyan,  Hrant and Iugov,  Vasilii and Mirumyan,  Mkhitar and Khachatryan,  Shahane and Hakobyan,  Tigran and Sedrakyan,  Tigran},
  year = {2023},
  month = dec 
}

@article{edge-z43,
  title = {Topological edge states in two-dimensional ${\mathbb{Z}}_{4}$ Potts paramagnet protected by the ${\mathbb{Z}}_{4}^{\ifmmode\times\else\texttimes\fi{}3}$ symmetry},
  author = {Topchyan, Hrant and Hakobyan, Tigran and Mirumyan, Mkhitar and Sedrakyan, Tigran A. and Sedrakyan, Ara},
  journal = {Phys. Rev. B},
  volume = {114},
  issue = {6},
  pages = {065111},
  numpages = {7},
  year = {2026},
  month = {Jul},
  publisher = {American Physical Society},
  doi = {10.1103/xszt-gsss},
  url = {https://link.aps.org/doi/10.1103/xszt-gsss}
}

@article{edge-z2f,
  title = {SPT extension of $Z_2$ quantum Ising model’s ferromagnetic phase},
  volume = {517},
  ISSN = {0375-9601},
  url = {http://dx.doi.org/10.1016/j.physleta.2024.129669},
  DOI = {10.1016/j.physleta.2024.129669},
  journal = {Physics Letters A},
  publisher = {Elsevier BV},
  author = {Topchyan,  Hrant},
  year = {2024},
  month = aug,
  pages = {129669}
}

@article{edge-z3,
	title={Two-dimensional topological paramagnets protected by $\mathbb{Z}_3$ symmetry: Properties of the boundary Hamiltonian},
	author={Hrant Topchyan and Vasilii Iugov and Mkhitar Mirumyan and Tigran Hakobyan and Tigran A. Sedrakyan and Ara G. Sedrakyan},
	journal={SciPost Phys.},
	volume={18},
	pages={068},
	year={2025},
	publisher={SciPost},
	doi={10.21468/SciPostPhys.18.2.068},
	url={https://scipost.org/10.21468/SciPostPhys.18.2.068},
}

@article{cohoms-math,
  title = {Classification of symmetry enriched topological phases with exactly solvable models},
  author = {Mesaros, Andrej and Ran, Ying},
  journal = {Phys. Rev. B},
  volume = {87},
  issue = {15},
  pages = {155115},
  numpages = {48},
  year = {2013},
  month = {Apr},
  publisher = {American Physical Society},
  doi = {10.1103/PhysRevB.87.155115},
  url = {https://link.aps.org/doi/10.1103/PhysRevB.87.155115}
}

@misc{thooft-gen-1,
  doi = {10.48550/ARXIV.1403.1467},
  url = {https://arxiv.org/abs/1403.1467},
  author = {Kapustin,  Anton},
  title = {Symmetry Protected Topological Phases,  Anomalies,  and Cobordisms: Beyond Group Cohomology},
  publisher = {arXiv},
  year = {2014},
  copyright = {arXiv.org perpetual,  non-exclusive license}
}

@article{thooft-gen-2,
  title = {Anomalous symmetries end at the boundary},
  volume = {2021},
  ISSN = {1029-8479},
  url = {http://dx.doi.org/10.1007/JHEP09(2021)017},
  DOI = {10.1007/jhep09(2021)017},
  number = {9},
  journal = {Journal of High Energy Physics},
  publisher = {Springer Science and Business Media LLC},
  author = {Thorngren,  Ryan and Wang,  Yifan},
  year = {2021},
  month = sep 
}

@article{thooft-finite,
  title = {Anomaly in Open Quantum Systems and its Implications on Mixed-State Quantum Phases},
  author = {Wang, Zijian and Li, Linhao},
  journal = {PRX Quantum},
  volume = {6},
  issue = {1},
  pages = {010347},
  numpages = {27},
  year = {2025},
  month = {Mar},
  publisher = {American Physical Society},
  doi = {10.1103/PRXQuantum.6.010347},
  url = {https://link.aps.org/doi/10.1103/PRXQuantum.6.010347}
}

@article{thooft-twisted,
	title={{Decorated defect construction of gapless-SPT states}},
	author={Linhao Li and Masaki Oshikawa and Yunqin Zheng},
	journal={SciPost Phys.},
	volume={17},
	pages={013},
	year={2024},
	publisher={SciPost},
	doi={10.21468/SciPostPhys.17.1.013},
	url={https://scipost.org/10.21468/SciPostPhys.17.1.013},
}

@article{continuum,
  title = {Continuum limit of spin-1 chain},
  author = {Inami, Takeo and Odake, Satoru},
  journal = {Phys. Rev. Lett.},
  volume = {70},
  issue = {13},
  pages = {2016--2019},
  numpages = {0},
  year = {1993},
  month = {Mar},
  publisher = {American Physical Society},
  doi = {10.1103/PhysRevLett.70.2016},
  url = {https://link.aps.org/doi/10.1103/PhysRevLett.70.2016}
}

@article{TL-orig,
  title = {Relations between the ‘percolation’ and ‘colouring’ problem and other graph-theoretical problems associated with regular planar lattices: some exact results for the ‘percolation’ problem},
  volume = {322},
  ISSN = {2053-9169},
  url = {http://dx.doi.org/10.1098/rspa.1971.0067},
  DOI = {10.1098/rspa.1971.0067},
  number = {1549},
  journal = {Proceedings of the Royal Society of London. A. Mathematical and Physical Sciences},
  publisher = {The Royal Society},
  author = {Temperley,  H. N. V. and Lieb,  Elliott H},
  year = {1971},
  month = apr,
  pages = {251–280}
}

@article{TL-integr-loop,
  title = {Bethe ansatz for the Temperley–Lieb loop model with open boundaries},
  volume = {2004},
  ISSN = {1742-5468},
  url = {http://dx.doi.org/10.1088/1742-5468/2004/03/P002},
  DOI = {10.1088/1742-5468/2004/03/p002},
  number = {03},
  journal = {Journal of Statistical Mechanics: Theory and Experiment},
  publisher = {IOP Publishing},
  author = {Gier,  Jan de and Pyatov,  Pavel},
  year = {2004},
  month = mar,
  pages = {P002–P002}
}

@article{TL-loop-ferm,
  title = {A Temperley–Lieb quantum chain with two- and three-site interactions},
  volume = {42},
  ISSN = {1751-8121},
  url = {http://dx.doi.org/10.1088/1751-8113/42/29/292002},
  DOI = {10.1088/1751-8113/42/29/292002},
  number = {29},
  journal = {Journal of Physics A: Mathematical and Theoretical},
  publisher = {IOP Publishing},
  author = {Ikhlef,  Y and Jacobsen,  J L and Saleur,  H},
  year = {2009},
  month = jul,
  pages = {292002}
}

@Article{TL-loop,
	title={{Fusion in the periodic Temperley-Lieb algebra and connectivity operators of loop models}},
	author={Yacine Ikhlef and Alexi Morin-Duchesne},
	journal={SciPost Phys.},
	volume={12},
	pages={030},
	year={2022},
	publisher={SciPost},
	doi={10.21468/SciPostPhys.12.1.030},
	url={https://scipost.org/10.21468/SciPostPhys.12.1.030},
}

@article{TL-rsos,
  title = {Factorization of density matrices in the critical RSOS models},
  volume = {2023},
  ISSN = {1742-5468},
  url = {http://dx.doi.org/10.1088/1742-5468/aceeef},
  DOI = {10.1088/1742-5468/aceeef},
  number = {8},
  journal = {Journal of Statistical Mechanics: Theory and Experiment},
  publisher = {IOP Publishing},
  author = {Westerfeld,  Daniel and Großpietsch,  Maxime and Kakuschke,  Hannes and Frahm,  Holger},
  year = {2023},
  month = aug,
  pages = {083104}
}

@article{cft-1,
  title = {Integrable lattice realizations of conformal twisted boundary conditions},
  volume = {517},
  ISSN = {0370-2693},
  url = {http://dx.doi.org/10.1016/S0370-2693(01)00982-0},
  DOI = {10.1016/s0370-2693(01)00982-0},
  number = {3–4},
  journal = {Physics Letters B},
  publisher = {Elsevier BV},
  author = {Chui,  C.H.Otto and Mercat,  Christian and Orrick,  William P. and Pearce,  Paul A.},
  year = {2001},
  month = oct,
  pages = {429–435}
}

@article{cft-2,
  title = {Conformal Field Theory from Lattice Fermions},
  volume = {398},
  ISSN = {1432-0916},
  url = {http://dx.doi.org/10.1007/s00220-022-04521-8},
  DOI = {10.1007/s00220-022-04521-8},
  number = {1},
  journal = {Communications in Mathematical Physics},
  publisher = {Springer Science and Business Media LLC},
  author = {Osborne,  Tobias J. and Stottmeister,  Alexander},
  year = {2022},
  month = nov,
  pages = {219–289}
}

@article{fuss-catalan,
  title = {Algebras associated to intermediate subfactors},
  volume = {128},
  ISSN = {1432-1297},
  url = {http://dx.doi.org/10.1007/s002220050137},
  DOI = {10.1007/s002220050137},
  number = {1},
  journal = {Inventiones Mathematicae},
  publisher = {Springer Science and Business Media LLC},
  author = {Jones,  Vaughan and Bisch,  Dietmar},
  year = {1997},
  month = Mar,
  pages = {89–157}
}

@article{integr,
title = {New integrable lattice models from Fuss-Catalan algebras},
journal = {Nuclear Physics B},
volume = {532},
number = {3},
pages = {609-634},
year = {1998},
issn = {0550-3213},
doi = {https://doi.org/10.1016/S0550-3213(98)00603-8},
url = {https://www.sciencedirect.com/science/article/pii/S0550321398006038},
author = {P. {Di Francesco}},
}

\end{document}